\documentclass[a4paper,fleqn,usenatbib]{mnras}

\usepackage{newtxtext,newtxmath}

\usepackage[T1]{fontenc}
\usepackage{ae,aecompl}

\usepackage{graphicx}	
\usepackage{amsmath}	
\usepackage{amssymb}	
\usepackage{enumitem}

\title[AGN Selection in IFU Surveys]{SDSS-IV MaNGA: Identification of active galactic nuclei in optical integral field unit surveys}

\author[Dominika Wylezalek]{Dominika Wylezalek$^{1}$\thanks{E-mail: dwylezalek@jhu.edu}, 
Nadia L. Zakamska$^{1}$,
Jenny E. Greene$^{2}$,
Rogemar A. Riffel$^{3,4}$,
\newauthor
Niv Drory$^{5}$,
Brett H. Andrews$^{6}$,
Andrea Merloni$^{7}$, 
Daniel Thomas$^{8}$
\\
$^{1}$ Department of Physics \& Astronomy, Johns Hopkins University, Bloomberg Center, 3400 N. Charles St., Baltimore, MD 21218, USA\\
$^{2}$ Department of Astrophysical Sciences, Princeton University, Princeton, NJ 08544, USA\\
$^{3}$ Departamento de F\'isica, CCNE, Universidade Federal de Santa Maria, Av. Roraima, 1000 - 97105-900, Santa Maria, RS, Brazil \\
$^{4}$ Laborat\'orio Interinstitucional de e-Astronomia - LIneA, Rua Gal. Jos\'e Cristino 77, Rio de Janeiro, RJ - 20921-400, Brazil\\
$^{5}$ McDonald Observatory, The University of Texas at Austin, 2515 Speedway Stop C1402, Austin, TX 78712, USA. \\
$^{6}$ PITT PACC, Department of Physics and Astronomy, University of Pittsburgh, Pittsburgh, PA 15260, USA\\
$^{7}$ Max-Planck-Institut f\"{u}r Extraterrestrische Physik, Giessenbachstra{\ss}e, 85748 Garching, Germany\\
$^{8}$ Institute of Cosmology and Gravitation, University of Portsmouth, Dennis Sciama Building, Portsmouth, PO1 3FX, UK\\
}

\date{Accepted XXX. Received YYY; in original form ZZZ}

\pubyear{2016}

\begin{document}
\label{firstpage}
\pagerange{\pageref{firstpage}--\pageref{lastpage}}
\maketitle

\begin{abstract}
In this paper, we investigate 2727 galaxies observed by MaNGA as of June 2016 to develop spatially resolved techniques for identifying signatures of active galactic nuclei (AGN). We identify 303 AGN candidates. The additional spatial dimension imposes challenges in identifying AGN due to contamination from diffuse ionized gas, extra-planar gas and photoionization by hot stars. We show that the combination of spatially-resolved line diagnostic diagrams and additional cuts on H$\alpha$ surface brightness and H$\alpha$ equivalent width can distinguish between AGN-like signatures and high-metallicity galaxies with LINER-like spectra. Low mass galaxies with high specific star formation rates are particularly difficult to diagnose and routinely show diagnostic line ratios outside of the standard star-formation locus. We develop a new diagnostic -- the distance from the standard diagnostic line in the line-ratios space -- to evaluate the significance of the deviation from the star-formation locus. We find 173 galaxies that would not have been selected as AGN candidates based on single-fibre spectral measurements but exhibit photoionization signatures suggestive of AGN activity in the MaNGA resolved observations, underscoring the power of large integral field unit (IFU) surveys. A complete census of these new AGN candidates is necessary to understand their nature and probe the complex co-evolution of supermassive black holes and their hosts.



\end{abstract}




\section{Introduction}

Most present-day massive galaxies host a supermassive black hole in their centers \citep{Magorrian_1998, Gebhardt_2000, Ferrarese_2000}, with $\sim 10$\% of them currently experiencing accretion activity powerful enough to be detected through optical or X-ray diagnostics \citep{Hao_2005a, Hao_2005, Kauffmann_2004}. Local relatively low-luminosity active galactic nuclei (AGNs) are historically called Seyfert galaxies \citep{Seyfert_1943}, while more distant and more powerful AGNs are historically called quasars, but it is now understood that at every epoch, the luminosity function of AGNs is a continuous function extending over many orders of magnitude \citep{Richards_2006b}. 

AGN have become a major and important component in modern galaxy formation models and theories. Both theoretical models and observations suggests that black holes and their energy output play an important role in shaping modern day galaxies \citep[e.g.][]{Croton_2006, Hopkins_2006, Somerville_2008, Novak_2011, Genel_2014, Alatalo_2014, Choi_2015, Smethurst_2015, Remus_2016}. But many of these details are still an active field of research. For example, the demographics of AGNs as a function of redshift and luminosity remains an important issue because the obscuration fraction is at the heart of many questions in black hole astrophysics and the obscuring material plays a key role in black hole evolution. Additionally, recent work on accretion timescales suggests that AGN flicker on short timescales of $\sim10^5$~yr, rather than remaining at a constant luminosity \citep{Schawinski_2015, Sun_2017} and may change type constantly as they increase or decrease in luminosity \citep{Elitzur_2014}. Such changing-look AGN may help to understand the accretion physics and associated timescales.

Much of this work relies on finding and identifying AGN in the first place. Seyfert galaxies and quasars present an enormous variety of observational signatures \citep{Antonucci_1993, Urry_1995}. In optical spectroscopic data, objects that appear with strong blue point-like continua and broad permitted emission lines (with velocity widths $\ga 1000$ km/sec) are termed type 1 AGNs, whereas objects with host galaxy-dominated continuum, no broad emission lines and high-ionization forbidden emission lines (with velocity widths of a few hundred km/sec) are termed type 2s. One of the early successes in understanding the structure of AGNs was the geometric unification model which explains these differences as being largely due to the orientation effects \citep{Antonucci_1985, Antonucci_1993}. In type 1 AGNs, the observer can see all the way down to the accretion disk around the supermassive black hole, whereas in type 2s, the observer's view is blocked by intervening clouds of gas and dust. As a result, the only optical indicator of AGN activity are the forbidden lines produced in extended material -- well outside of the obscuring region --  which is illuminated and photo-ionized by the hidden nucleus. 

Most AGN identification techniques in optical spectroscopic surveys therefore rely on identifying AGN via their illumination of the gas. Emission line flux ratios and diagnostic diagrams are the most common way to identify AGN in optical spectra \citep{Baldwin_1981, Veilleux_1987, Zakamska_2003, Hao_2005a, Kauffmann_2003, Kewley_2006, Reyes_2008, Yuan_2016}. But a major caveat of large optical spectroscopic surveys such as the Sloan Digital Sky Survey \citep[SDSS;][]{York_2000, Gunn_2006} is the small size of the optical fibres which, at  3" diameter (in the case of SDSS-I to SDSS-III surveys), only cover a tiny fraction of the footprint of a galaxy and are only sensitive to processes close to the galactic center. \citet{Wylezalek_2017} has recently shown that AGN activity can easily be hidden in the integrated spectrum of the galaxy. This may happen if the AGN-ionized regions are predominantly in the outskirts of the galaxy due to obscuration, other ionization processes dominate in the center of the galaxy, or if simply the AGN is offset from the galaxy center due to a recent merger \citep{Greene_2011, Comerford_2012, Comerford_2014, Hainline_2016}. Another possibility is that the AGN has recently turned off \citep{Shapovalova_2010, McElroy_2016} and relic ionization signatures, so called light echoes, are only visible at large distances from the center \citep{Keel_2012, Keel_2015}. Much of such AGN activity might therefore have been missed or underestimated in the past with potentially significant implications for AGN and galaxy evolution models. 

Integral field unit (IFU) surveys now offer new possibilities in finding such `hidden' spatially extended AGN signatures. The SDSS-IV \citep{Blanton_2017} survey Mapping Nearby Galaxies at APO \citep[MaNGA;][]{Bundy_2015, Drory_2015, Law_2015, Yan_2016a, Yan_2016b} is a new optical fibre-bundle IFU survey and will obtain IFU observations of 10,000 galaxies at $z \lesssim 0.1$ over the next few years, allowing an extensive investigation of the spatial dimension of galaxy evolution. MaNGA will be particularly crucial for AGN science since it may provide us with a completely new census of AGN activity in the nearby Universe. Additionally, through a large sample of spatially resolved spectroscopic observations of AGN where simultaneous and spatially resolved measurements of different gas phases and stellar activity can be obtained, the survey also has the potential to make major contributions to our understanding of feeding and feedback processes in low- and intermediate luminosity AGN \citep{Wylezalek_2017}. To achieve ambitious science goals, it is crucial to develop a robust selection mechanism for AGN that not only allows to uncover the kind of AGN activity that is already known about but also leaves room for much of the `unexpected' and `weird' AGN activity that might have been previously missed. At the same time, such a selection also needs to address the problems and challenges that come with finding and characterizing AGN signatures at large galaxy distances, particularly crucial when dealing with large IFU surveys \citep{Belfiore_2016, Jones_2017, Zhang_2017}.

In this paper, which is based on a quarter of the final MaNGA sample, we provide a census of AGN and AGN-like signatures as obtained by exclusively using the spatially resolved MaNGA spectra. We develop AGN selection criteria for IFU surveys like MaNGA that are based on identifying AGN via their illumination of the gas. In a forthcoming paper, we will explore the multi-wavelength nature of the AGN selected based on this algorithm and we will also study their effects on the gas kinematics, as is being done in studies of galaxy-wide feedback. 

The paper is organized as follows. In Section 2 we present the data and emission line measurements. In Section 3, we present our initial AGN selection criterion based on selection methods in the literature and show that they are not sufficient for AGN selection in IFU surveys. In Section 4 we then present a refined AGN selection criteria optimized for the MaNGA survey and compare our AGN candidates to results from SDSS-III single fibre observations and previous AGN selections in Section 5. Section 6 concludes our findings. Throughout the paper we use $H_0 = 72$ km s$^{-1}$ Mpc$^{-1}$, $\Omega_m = 0.3, \Omega_{\Lambda} = 0.7$.

\section{Data}

\subsection{The MaNGA Survey}

The MaNGA survey is one of the three major parts of the ongoing 4th phase of the SDSS (SDSS-IV) and is an optical fibre-bundle IFU spectroscopic survey. MaNGA utilizes the Baryon Oscillation Spectroscopic Survey (BOSS) spectrograph \citep{Smee_2013} with a spectral coverage of $3622-10354$~\AA\ at $R \sim 2000$.The fibres have a size of 2\arcsec\ aperture (2.5\arcsec\ separation between fibre centers), which at $z\sim 0.05$ corresponds to $\sim 2$ kpc, although with dithering the effective sampling improves to $1.4$\arcsec. The bundle sizes vary between 19 and 127 (or 12.5$-$32.5\arcsec diameter on the sky) depending on the apparent size of the target galaxy, leading to an average footprint of $400-500$~arcsec$^{2}$ per IFU. Between 2014 and 2020, MaNGA will have obtained IFU observations of about 10,000 galaxies at $z\la 0.1$ and with stellar masses $>10^9 M_{\odot}$, spanning a wide range of environments, thereby allowing an extensive investigation of the spatial dimension of galaxy evolution \citep{Bundy_2015, Yan_2016a}. Targets are selected based on SDSS-I redshifts and $i$-band luminosity to achieve uniform radial coverage in terms of the effective radius $R_e$ and a nearly flat distribution in stellar mass. MaNGA will observe 2/3 of the sample out to $\sim 1.5 R_e$ and 1/3 of the sample out to $\sim 2.5 R_e$.  

\subsection{MaNGA Data Products}

All analysis in this paper is based on the fifth internal MaNGA Product Launch which was released to the collaboration on June 27 2016 and consists of data cubes for a total of 338 plates, corresponding to 2778 galaxies of which 2727 are unique galaxies. The MaNGA Data Reduction Pipeline produces sky-subtracted spectrophotometrically calibrated spectra and rectified three-dimensional data cubes that combine individual dithered observations \citep[for details on MaNGA data reduction see][]{Law_2016}. Relative flux calibration for the MaNGA data is better than 5\% and is described by \citet{Yan_2016b}. 
The MaNGA Data Analysis Pipeline \citep[][]{Westfall_2017} is a project-led software package used to analyze the data products provided by the MaNGA Data Reduction Pipeline with the goal of providing the collaboration and public with survey-level quantities, such as kinematics, emission-line properties, and stellar-population parameters. In addition to analysing each individual spaxel, the Data Analysis Pipeline constructs spatially binned data cubes (totally, radially binned from $0-1$ and $1-2$ $R_{eff}$ and Voronoi-binned cubes) and performs data analysis on these binned spectra. The MaNGA Data Analysis Pipeline first fits the stellar continuum using the Penalized Pixel-Fitting method \citep[pPXF, ][]{Cappellari_2004, Cappellari_2017} and then subtracts the best-fitting stellar continuum from the observed data using the MILES\footnote{http://www.iac.es/proyecto/miles/pages/stellar-libraries/miles-library.php} stellar templates before fitting the emission lines. The fit to the emission lines allows for a non-zero baseline. The main output products of the Data Analysis Pipeline then include the stellar absorption line kinematics (stellar velocity and stellar velocity dispersion measurements) and emission line measurements of 21 major optical emission lines in the MaNGA spectral range. Both non-parametric parameters (emission line flux, equivalent width) and Gaussian-profile measurements (emission line flux, velocity, velocity dispersion) are provided \citep[see also][]{Cherinka_2017}. We note that the Gaussian profile velocity dispersion $\sigma_{line}$ still needs to be corrected for the instrumental dispersion $\sigma_{inst}$ at the fitted line center by the user through $\sigma_{line,corr} = \sqrt{\sigma_{line}^2 - \sigma_{inst}^2}$. The instrumental dispersion is an output product of the Data Analysis Pipeline. 

In this paper, we utilize the line kinematic measurements provided by the MaNGA Data Analysis Pipeline. Specifically, we use the Data Analysis Pipeline products of type {\tt SPX-GAU-MILESHC} which contain analysis results of each individual pixel and which are geared towards emission-line science. The stellar continuum is only fit to spectra with a $r$-band signal-to-noise $S/N > 1$.

\subsection{Other Data Products}

In addition to the measurements obtained directly from the MaNGA data set, we make use of added-value products from the NASA-Sloan Atlas (NSA catalog\footnote{http://nsatlas.org}). The NSA catalog is a catalog of images and parameters derived from SDSS imaging, and with the addition of Galaxy Evolution Explorer (GALEX) data for the ultraviolet part of the spectrum \citep{Blanton_2011}. In the analysis of this paper, we use the NSA-derived redshifts, Sersic stellar masses and the $r$-band based effective radii.

\subsection{MaNGA AGN Ancillary Sample}

In addition to the main MaNGA sample, a dedicated MaNGA-AGN program (PI: J.E.Greene) was awarded 120 MaNGA-IFU observations. The goal of the AGN Ancillary Program is to extend the luminosity range of AGN in the main MaNGA sample to a bolometric luminosity of $L_{\rm{bol}} \sim 10^{45}$~erg/s. The sample consists of AGN that were selected at different wavelengths. The first subsample is composed of AGN selected based on their hard X-ray emission from the the Burst Alert Telescope (BAT) aboard the \textit{Swift} satellite \citep{Ajello_2012}. The second sample is selected from the AGN Line Profile And Kinematics Archive \citep[ALPAKA, ][]{Mullaney_2013} which is mainly based on [OIII]$5007\AA$ line flux measurements. The third sample is a mid-IR selected sample based on observations with the \textit{Wide-field Infrared Survey Explorer} \citep[WISE][]{Wright_2010} satellite with $0.7 < W1-W2 <2.0$; $2.0<W2-W3<4.5$, following the color-selection prescriptions in \citet{Wright_2010}. These three samples are matched in bolometric luminosity and redshift, keeping the [OIII] and WISE samples within $0.01 < z < 0.08$ so that a roughly comparable spatial coverage of all of them can be obtained. Thus, the redshift range of the AGN sample is somewhat broader than that of the MaNGA main sample.

As of June 27 2016, the fifth internal MaNGA data release date, 13 ancillary AGN had been observed by MaNGA. These include 1 BAT-selected AGN \citep[][]{Wylezalek_2017}, 7 [OIII]-selected AGN and 5 WISE-selected AGN (Table \ref{ancillary}).

\begin{table*}
\caption{Source Information of the MaNGA Ancillary AGN}
\begin{center}
\begin{tabular}{lcccccc}
\hline\hline
MaNGA ID & MaNGA plate-ID & MaNGA IFU-ID & R.A. & Dec. & z &Selection\\
\hline\vspace{0.05cm}

1-137883&                   8249& 3704 &       137.87476 &       45.468320 &     0.026825300& BAT \\
1-209980 &                  8549& 12701&        240.47087 &       45.351940 &     0.042046800& [OIII]  \\
1-44303 &                  8718& 12701&        119.18215 &       44.856709 &     0.049920000& [OIII] \\
1-593159 &                  8547& 12701&        217.62997 &       52.707159 &     0.044881100& [OIII] \\
1-44379 &                  8718& 12702&        120.70071 &       45.034554 &     0.038928000& [OIII] \\
1-96151 &                  8612& 12704&        254.56457 &       39.391464 &     0.034311600& [OIII] \\
1-339094 &                  8141& 1901&        117.47242 &       45.248483 &     0.031259100& [OIII] \\
1-23979 &                  7991& 3702&        258.15875 &       57.322421 &     0.026629800& [OIII] \\
1-47256 &                  8724 &12702&        132.01706 &       54.028023 &     0.049333900& WISE \\
1-177270 &                  8613 &12703&        256.81775 &       34.822605 &     0.036685800& WISE \\
1-24423 &                  8626 &12704&        263.75522 &       57.052433 &     0.047232300& WISE \\
1-90901 &                  8553 &12705&        235.46388 &       55.467854 &     0.048214000& WISE \\
1-149211 &                  8947 &3701&        168.94780 &       50.401634 &     0.047306800& WISE \\
\hline 
\end{tabular}
\end{center}
\label{ancillary}
\end{table*}

\section{Selecting AGN Candidates in MaNGA}
 
\subsection{Diagnostic Diagrams}

Selecting AGN based on broad Balmer emission lines is a powerful way of identifying type-1, i.e. unobscured, AGN. These broad emission lines have Doppler widths in the range $1000 - 25000$~km s$^{-1 }$ and originate in high density $n_e > 10^9$~cm$^{-3}$ gas close to the black hole \citep[][and references therein]{Netzer_2015}. In obscured, type-2 AGN, such broad emission lines are absent and AGN have to be separated from inactive galaxies using other methods. Type-2 AGN are characterized by strong emission lines, but so are star-forming galaxies. To distinguish between these populations, the most commonly applied method in optical spectroscopic surveys of low-redshift galaxies is utilizing diagnostic diagrams \citep{Baldwin_1981, Veilleux_1987, Kauffmann_2003, Kewley_2006}. The foundation of this method lies in measuring the ionization state of the gas through various emission line ratios, with high-ionization lines being indicative of a hard ultra-violet ionizing continuum which can only be produced in an AGN.

Diagnostic diagrams are typically constructed using a set of nebular emission line ratios which distinguish between different ionization mechanisms. The most commonly used ones are the BPT diagrams \citep{Baldwin_1981, Veilleux_1987} using [NII]6584/H$\alpha$ versus [OIII]5007/H$\beta$ ([NII]-BPT diagram), [SII]6717,6731/H$\alpha$ versus [OIII]5007/H$\beta$ ([SII]-BPT) and [OI]6300/H$\alpha$ versus [OIII]5007/H$\beta$ ([OI]-BPT). A major advantage of the BPT diagnostic diagrams is that the required emission lines are relatively close in wavelength space such that usually all of them can be observed in one optical spectrum. Furthermore, the most commonly used ratios are composed of lines so close together that the ratios are little affected by interstellar extinction. Depending on ionization models several dividing lines have been developed such that the diagrams can be used to distinguish between different ionization mechanisms such as ionization through star formation, AGN or shocks. In this paper, we use the dividing lines developed by \citet{Kewley_2001} and \citet{Kauffmann_2003} and summarized in \citet{Kewley_2006}. 

Specifically, the [SII]-BPT allows to distinguish between star-formation, AGN or `low ionisation nuclear emission line regions' \citep[LINER,][]{Heckman_1980} dominated emission line regions. The [NII]-BPT diagram allows to distinguish between star-formation, AGN/LINER, or Composite (mix of AGN and star formation) dominated emission line regions. Because the [NII]-BPT diagram does not separate well between AGN or LINER-like emission, in the remaining part of the paper we simply refer to this emission in the [NII]-BPT as `AGN'-like. LINER spectra show strong low ionization emission lines and characteristic line ratios which makes them more identifiable in the [SII]-BPT diagram than in the [NII]-BPT. 

LINER-like emission is associated with a number of ionization mechanisms: weakly ionizing AGN \citep{Heckman_1980}, shock ionization (either related to star-forming processes in inactive galaxies or AGN activity) or photo-ionization through hot evolved stars \citep[see e.g.][and references therein]{Binette_1994, Ho_2008, Stasinska_2008, Eracleous_2010, Sarzi_2010,  Yan_2012, Singh_2013, Belfiore_2016}. In nuclear spectra, there is strong evidence tying LINER lines to AGN activity \citep{Ho_2008} but even then it is not clear that the AGN provides the dominant power source \citep{Eracleous_2010}. Spatially resolved emission likely has another origin. In the era of increasing samples of galaxies with IFU observations, LINER-like emission is now frequently detected on galaxy-wide scales with no evidence for young stellar populations or AGN activity. Using a sample of $\sim 650$ galaxies observed with MaNGA (as of April 2015), \citet{Belfiore_2016} argue that post asymptotic giant branch (pAGB) stars can produce the required hard ionizing spectrum to power the emission in galaxies with extended LI(N)ER-like (leaving out the `N' for nuclear) emission line regions (eLIER galaxies) and also in galaxies that show centrally peaked LI(N)ER-like emission line region \citep[see also ][]{Binette_1994, Sarzi_2010}. In both cases, LIER emission line regions show no sign of young stellar populations and their emission-line flux follows that of the old stellar continuum as traced by low H$\alpha$ equivalent widths and by stellar population indices, respectively. Although the extremely hot component of an old stellar population is poorly understood \citep{OConnell_1999, Dotter_2007, Conroy_2013, Choi_2017}, it is thought that these stars cannot produce equivalent widths of H$\alpha$ in excess of $\sim 3$\AA. 

In addition to LI(N)ER emission being associated with either ionization through AGN, shock-ionization (related to AGN or star-forming processes) or ionization through old stars, \citet{Zhang_2017} have recently shown that regions in the galaxy that are dominated by diffuse ionized gas \citep[][]{Reynolds_1984, Dettmar_1990, Rossa_2003a, Rossa_2003b, Oey_2007} can also mimic LI(N)ER-like line ratios in the [SII]-BPT and even Composite and AGN-emission line ratios in the [NII]-BPT diagram \citep[see also][]{Reynolds_1985, Hoopes_2003, Madsen_2006}. This effect is related to changes of the ionization parameter $U$, metallicity $Z$ compared to star forming regions, i.e. H\,{\sevensize II}, regions. \citet{Zhang_2017} further suggest that a harder ionizing spectrum is contributing to enhancing the emission lines ratios in diffuse ionized gas regions which may be related to ionization by evolved stars. Therefore contribution from diffuse ionized gas regions can be particularly important in post-starburst galaxies and galaxies with old stellar populations. The contribution of diffuse ionized gas regions to the total emission line flux can be significant, in particular in face-on galaxies \citep{Oey_2007}. 

The phenomenon of increased emission line ratios in diffuse ionized gas regions and galaxies where ionization through pAGB stars mimic LI(N)ER-like line ratios may have a common origin, although there might be subtle differences in the stellar populations and metallicities in classical LI(N)ER-selected galaxies and diffuse ionized gas-selected galaxies \citep{Zhang_2017}. These studies show, however, that selecting AGN in resolved IFU observations on the basis of simple line-ratio diagnostics is not a straight forward task. The impact of AGN and extended LINER emission that is not associated with AGN becomes increasingly important in spatially resolved spectroscopic observations. 

In addition to the classical line ratio diagnostic diagrams, \citet{Cid-Fernandes_2010} have shown that invoking the equivalent width of the H$\alpha$ emission line $EW({H\alpha})$ allows to differentiate between the different ionization mechanisms that lead to the overlap in the LI(N)ER region of traditional diagnostic diagrams. Based on the bimodal distribution of $EW({H\alpha})$, \citet{Cid-Fernandes_2010} suggest that $EW({H\alpha}) > 3$\AA\ optimally divides true AGN from `fake' AGN in which LI(N)ER emission is due to hot evolved stars.  Based on these findings \citet{Cid-Fernandes_2011} advertise a new diagnostic diagram using $EW({H\alpha})$ vs. [NII]/H$\alpha$. In this diagnostic diagram, AGN are classified as sources with $EW({H\alpha}) > 3$~\AA\ and  [NII]/H$\alpha > -0.4$, while star-forming galaxies are classified as sources with $EW({H\alpha}) > 3$\AA\ and and  [NII]/H$\alpha < -0.4$. `Fake AGN' are sources with [NII]/H$\alpha > -0.4$ but $EW({H\alpha}) < 3$~\AA.

A large fraction of $\sim 70$~\% of MaNGA galaxies show strongly detected emission lines in the AGN, LINER or Composite regions of the diagnostic diagrams. As we discussed above, multiple possible sources of ionization need to be disentangled to identify AGN in large optical IFU surveys.

\subsection{Selection Algorithm}

Using the non-parametric emission line measurements provided by the MaNGA Data Analysis Pipeline, we utilize both the traditional [NII] and [SII]-BPT diagrams and construct resolved BPT-maps for all galaxies in MaNGA sample. To do so, we first exclude all spaxels with an $r$-band signal-to-noise ratio $S/N < 5$ and spaxels in which either the MaNGA Data Analysis Pipeline or MaNGA Data Reduction Pipeline failed to reconstruct the spectrum in that spaxel and/or failed to perform the data analysis tasks (i.e. stellar continuum fitting, emission line measurements). These spaxels are marked as `CRITICAL' in the {\tt MaNGA\_DAPQUAL} quality masks that are provided as part of the Data Analysis Pipeline product package for each emission line fit that was attempted. Spaxels that are excluded based on this S/N cut or spaxels that are flagged as `CRITICAL' in any relevant emission line map are disregarded from any further analysis. These spaxels are also not regarded when computing spaxel fractions (see below).

\begin{figure*}
\centering
\includegraphics[scale = 0.36, trim = 4cm 59cm 0cm 2.5cm, clip= true]{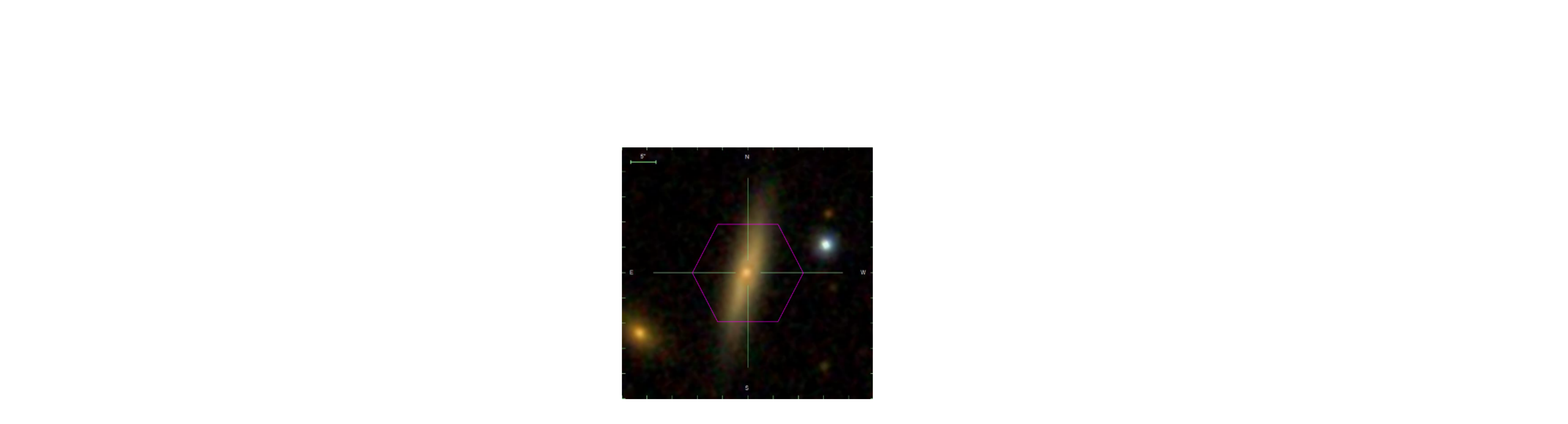}
\includegraphics[scale = 0.36, trim = 4cm 6cm 0cm 27.5cm, clip= true]{./plots/8138_6101.pdf}
\caption{Example classification plots for MaNGA:1-339010 $z = 0.03$. The upper-most panel shows the SDSS composite $gri$ optical image. The two left columns then show the results of the [NII] BPT analysis. We show the [NII]-resolved BPT, where AGN-like emission is colored in red, Composite-like emission is colored in orange and star-forming-like emission is colored in cyan. The left plot next to the resolved BPT shows the position of the emission-line measurements per spaxel in the [NII] BPT space. We then also show the the optical spectra obtained within MaNGA, zoomed into the relevant spectral regions with the H$\beta$, [OIII], H$\alpha$, [NII] and [SII] emission lines marked. The red spectrum shows the stacked spectrum in the red, i.e. AGN-like spaxels, the orange spectrum shows the stacked spectrum of the orange, i.e. Composite, spaxels and the blue spectrum shows the stacked spectrum of the blue, i.e. star-forming, spaxels. The black dashed spectrum shows the sum of the best fitted spectral model plus emission lines from the MaNGA Data Analysis Pipeline. We repeat the same kind of analysis for the [SII] BPT classification (left two columns), where orange spaxels refer to the AGN-like emission, pink spaxels to the LI(N)ER like emission and dark blue spaxels to star-forming-like emission.}
\label{BPT_example}
\end{figure*}

We then classify each spaxel based on its position in both the [NII]- and [SII]-BPT diagram independently. That means that a single spaxel may have one classification based on the [NII]-BPT, but a different classification based on the [SII]-BPT \citep[an example is discussed by][]{Wylezalek_2017}. We show an example of this classification in Figure \ref{BPT_example}. 

Based on these diagrams, we define the following parameters:

\begin{itemize}
\item f$_{\rm{A, N}}$: The \textit{[NII]-BPT AGN spaxel fraction}, i.e. the summed fraction of spaxels that are classified as `AGN' or `Composite' in the [NII]-BPT diagram. To account for the contribution of processes associated with star formation, spaxels that are classified as `Composite' are given a weight of 20\% and spaxels that are classified as `AGN' are given a weight of 80\% in the computation of f$_{\rm{AGN, [NII]}}$.
\item f$_{\rm{A, S}}$: The \textit{[SII]-BPT AGN spaxel fraction}, i.e. the fraction of spaxels that are classified as `AGN' in the [SII]-BPT diagram.
\item f$_{\rm{L, S}}$: The \textit{[SII]-BPT LIER spaxel fraction}, the fraction of spaxels that are classified as `LI(N)ER' in the [SII]-BPT diagram.
\item f$_{\rm{AL, S}}$: The \textit{combined [SII]-BPT AGN and LIER spaxel fraction}, i.e. f$_{\rm{A, S}} +$ f$_{\rm{L, S}}$
\item $EW(H\alpha)_{\rm{A, N}}$: The \textit{[NII]-BPT AGN EW(H$\alpha$) measure}, i.e. the mean value of the top 20\% percentile of the distribution of the equivalent widths of the H$\alpha$ emission line in spaxels that are classified as `AGN' or `Composite' in the [NII]-BPT diagram.
\item $EW(H\alpha)_{\rm{A, S}}$: The \textit{[SII]-BPT AGN EW(H$\alpha$) measure}, i.e. the mean value of the top 20\% percentile of the distributions of equivalent widths of the H$\alpha$ emission line in spaxels that are classified as `AGN' in the [SII]-BPT diagram.
\item $EW(H\alpha)_{\rm{L, S}}$: The \textit{[SII]-BPT LIER EW(H$\alpha$) measure}, i.e. the mean value of the top 20\% percentile of the distributions of equivalent widths of the H$\alpha$ emission line in spaxels that are classified as `LI(N)ER' in the [SII]-BPT diagram.
\item $EW(H\alpha)_{\rm{AL, S}}$: The \textit{[SII]-BPT AGN+LIER EW(H$\alpha$) measure}, i.e. the mean value of $EW(H\alpha)_{\rm{A, S}}$ and  $EW(H\alpha)_{\rm{L, S}}$.
\end{itemize} 

We then flag MaNGA-observed galaxies as AGN candidates if the source fulfills either of the two following conditions:

\begin{itemize}
\item f$_{\rm{A,N}}$ > 15\% and $EW(H\alpha)_{\rm{A, N}} > 5$ \AA

or

\item f$_{\rm{AL, S}}$ > 15\% and $EW(H\alpha)_{\rm{AL, S}} > 5$ \AA. 
\end{itemize}

The average signal-to-noise ratio of the emission line fluxes per spaxel are $9, 11, 11, 13$ and 46 for the H$\beta$, [OIII], [NII], [SII] and H$\alpha$ lines, respectively. These selection criteria are driven by several constraints: Using higher resolution optical IFU follow-up observations and multi-wavelength analysis of two MaNGA-selected AGN candidates, \citet{Wylezalek_2017} have shown that objects with only small AGN- oder LI(N)ER-like regions in the resolved [NII]- or [SII]-BPT maps can host low- and intermediate luminosity AGN. One of the objects studied by \citet{Wylezalek_2017} only showed LI(N)ER-like emission in the resolved [SII]-BPT map. This emission is most likely due to shock excitation due to AGN-driven outflows propagating through the interstellar medium or small-scale radio jets that inflate over-pressured cocoons providing shock ionization \citep{Croston_2007}. This example shows that galaxies with small combined [SII]-BPT AGN and LIER spaxel fractions f$_{\rm{AL, S}}$ or small [NII]-BPT AGN spaxel fractions f$_{\rm{A,N}}$ may host low- to intermediate-luminosity AGN that are either too low-luminosity or perhaps too obscured over a large covering factor so that their ionization signatures are not seen over large scales \citep[see also ][]{Hainline_2014, Lena_2015, Fischer_2016}, hence our choice of relatively small f$_{\rm{AL, S}}$ and f$_{\rm{A, N}}$. 

Our selection criteria are furthermore driven by the models and observations of emission-line ratios in diffuse ionized gas regions and due to hot evolved stars. As mentioned above, both phenomena can lead to enhanced line ratios that mimic LI(N)ER-like emission. We therefore employ the additional cut on the H$\alpha$ equivalent width in AGN and LI(N)ER-dominated spaxels. Equivalent widths of $\la 3$\AA\ are compatible with the expectations from reprocessing of the ionising radiation from hot evolved stars \citep{Binette_1994, Stasinska_2008, Cid-Fernandes_2011}. To compute the equivalent width measures $EW(H\alpha)_{\rm{A, N}}$, $EW(H\alpha)_{\rm{A, S}}$ and $EW(H\alpha)_{\rm{L, S}}$, we only choose the top 20th percentile of the distribution of H$\alpha$ equivalent widths in AGN and LI(N)ER-classified spaxels, respectively. This choice is motivated by the frequent occurrence of signatures of diffuse ionized gas regions, hot stars, AGN and shocks in a single galaxy. If, for example, emission-line diagnostics are dominated by ionization through AGN activity in the center of a galaxy, but contributions of diffuse ionized gas regions or old stars dominate the outskirts of a galaxy, then the total footprint of the galaxy may be classified as AGN or LI(N)ER-like. But due to the different nature of the ionization mechanisms in different parts of the galaxy, the mean H$\alpha$ equivalent width in all AGN+LI(N)ER-dominated spaxels may not pass the typically applied threshold of $5$~\AA. An example for such a source is discussed in Section 5.2. To not exclude bona-fide AGN such as this source, we implement the cut in $EW(H\alpha)_{\rm{A, N}}$, $EW(H\alpha)_{\rm{A, S}}$ and $EW(H\alpha)_{\rm{L, S}}$ as described above.

This selection leaves us with 746 galaxies, of which 733 are unique galaxies. We refer to this sample as the `initial AGN candidates'.

\subsection{Spaxel Fraction Measurements}

\begin{figure*}
\centering
\includegraphics[scale = 0.485, trim = 0.8cm 0cm 1.7cm 1cm, clip= true]{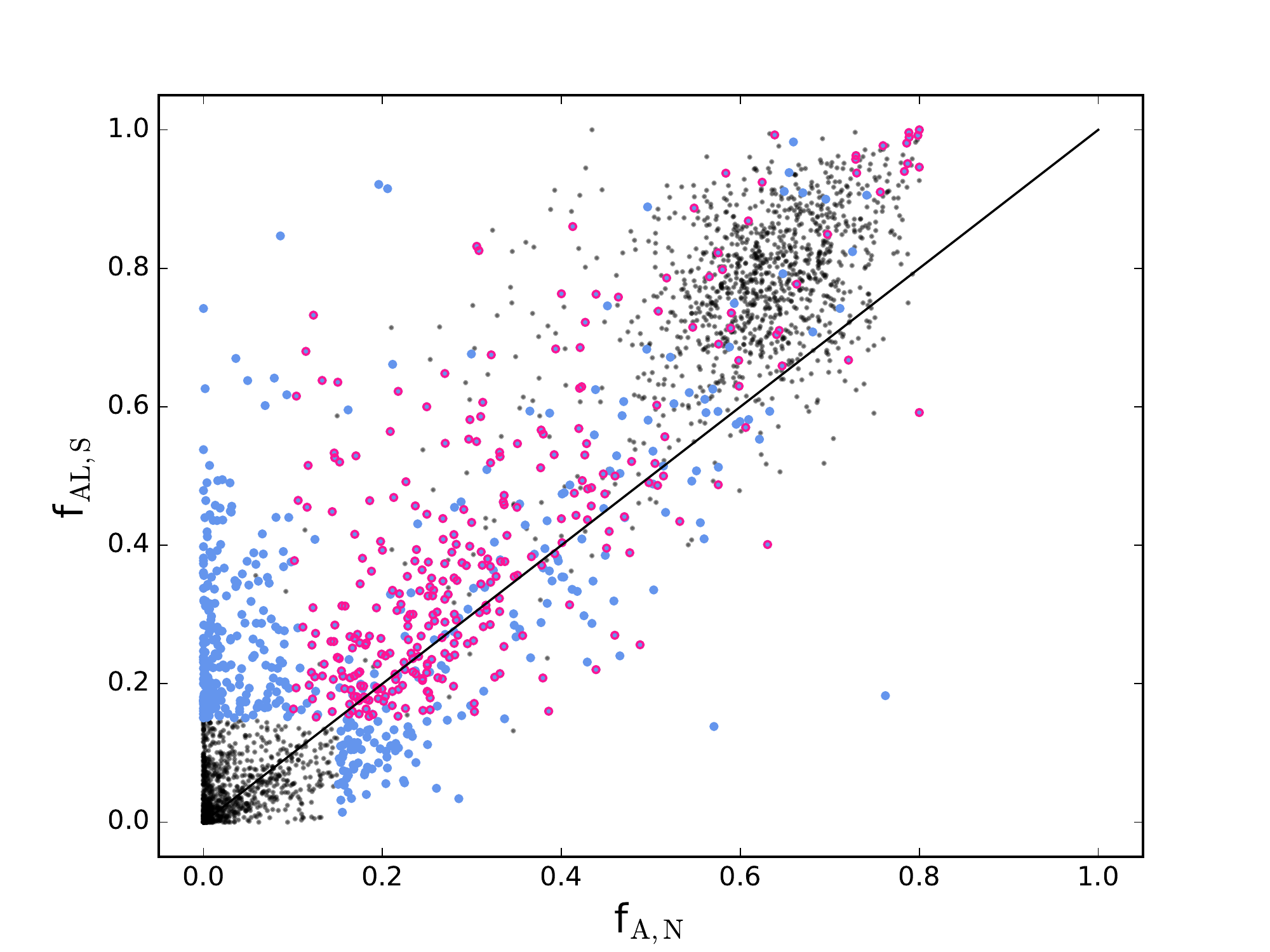}
\includegraphics[scale = 0.485, trim = 0.8cm 0cm 1.7cm 1cm, clip= true]{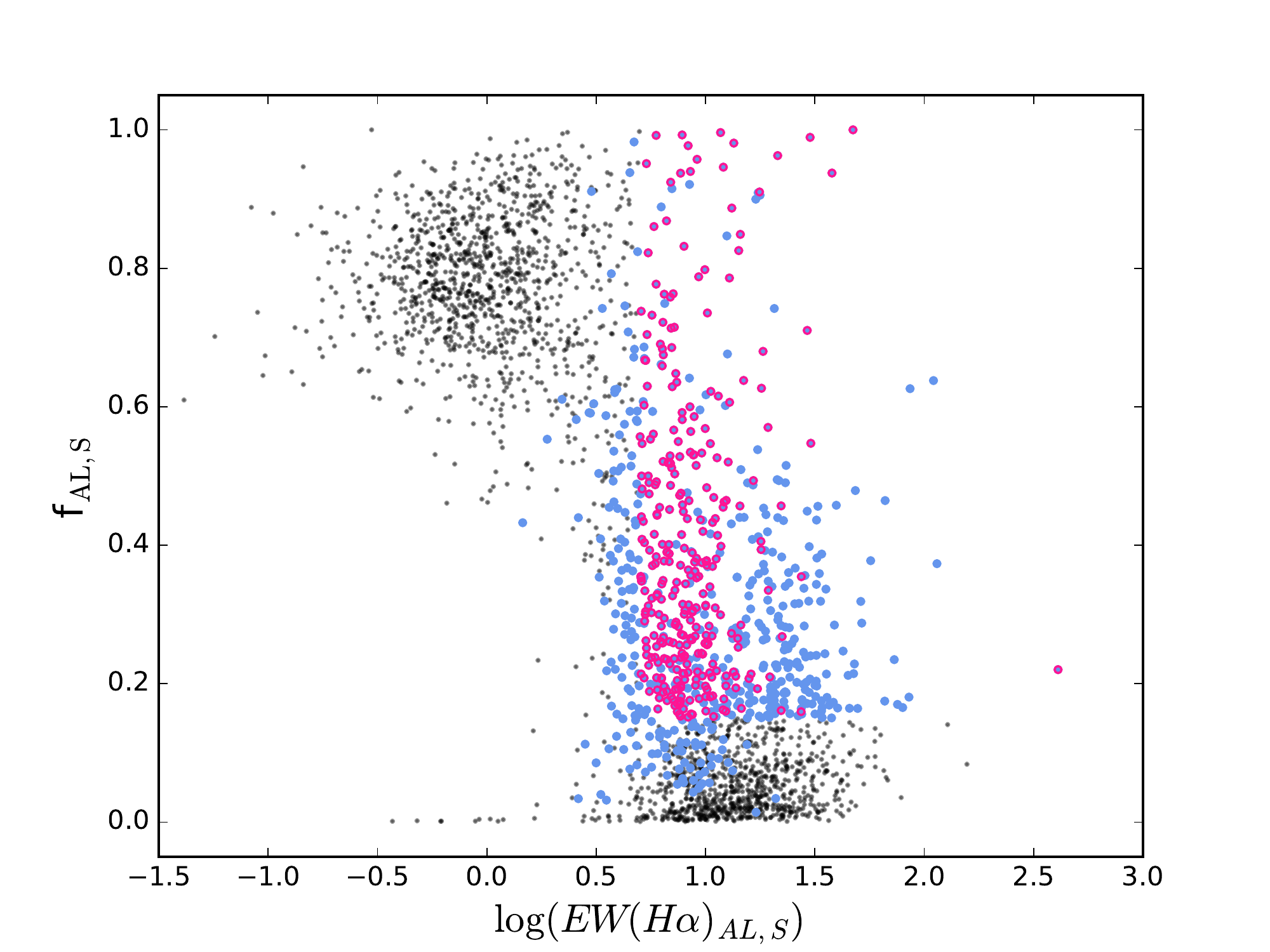}
\caption{\textbf{Left:}  The [NII]-BPT based AGN spaxel fraction f$_{\rm{A, N}}$ vs. the combined [SII]-BPT AGN and LIER spaxel fraction f$_{\rm{AL, S}}$ for the whole MaNGA sample (black, small data points) and for the initial 746 AGN candidates selected in this paper (blue large data points). The whole MaNGA sample generally shows a low spaxel fraction and a high spaxel fraction population. Only few AGN candidates are selected from the high spaxel fraction population. This is due to their the low $EW(H\alpha)$ in the relevant spaxels which we show in the \textbf{right} panel: Most sources with high spaxel fractions in the whole MaNGA sample (small black points) have $EW(H\alpha) < 5$\AA\ (marked by the vertical dashed line) and are therefore not selected as AGN candidates which are shown in blue.}
\label{fractions}
\end{figure*}

In the left panel of Figure \ref{fractions} we show how the AGN spaxel fractions as determined from the two different BPT diagrams, f$_{\rm{A, N}}$ and f$_{\rm{AL, S}}$, compare with each other for all galaxies in the MaNGA sample (black small data points) and for the selected sources (blue data points). For the whole sample (small, black data points), generally two large populations can be seen: one with high spaxel fraction and one with low spaxel fraction. In the high spaxel fraction population, both the [NII]-BPT based AGN spaxel fraction f$_{\rm{A, N}}$ and the combined [SII]-BPT AGN and LIER spaxel fraction f$_{\rm{AL, S}}$ are $> 0.5$, while f$_{\rm{AL, S}}$ generally tends to be somewhat higher than f$_{\rm{A, N}}$. In the low spaxel fraction population, the [NII]-BPT based AGN spaxel fraction f$_{\rm{A, N}}$ and the combined [SII]-BPT AGN and LIER spaxel fraction f$_{\rm{AL, S}}$ are  $<  0.3$. 

Focusing on the sources that the AGN detection algorithm does select (i.e. blue, larger points Figure \ref{fractions}), we notice two sub-populations where the scatter between the [NII]-BPT based AGN spaxel fraction f$_{\rm{A, N}}$ and the combined [SII]-BPT AGN and LIER spaxel fraction f$_{\rm{AL, S}}$ is large. Some sources can show large values of f$_{\rm{A, N}}$ while having low f$_{\rm{AL, S}}$. This behaviour is not necessarily surprising, since the [NII]-BPT based AGN spaxel fraction f$_{\rm{A, N}}$ is computed as a weighted summed fraction of `AGN' or `Composite' classified in the [NII]-BPT diagram. The `Composite' classification is more inclusive than the 'AGN' and 'LIER' classifications in the [SII]-BPT diagram, which is why a single galaxy may be completely dominated by `Composite' emission in the resolved [NII]-BPT and therefore have a large f$_{\rm{A, N}}$, while the [SII]-BPT classification would identify the spaxels/galaxy as `star-forming', leading to a low f$_{\rm{AL, S}}$. 

However, when large [NII]-BPT based AGN spaxel fractions f$_{\rm{A, N}}$ coupled with high [NII]-BPT AGN EW(H$\alpha$) measures of $EW(H\alpha)_{\rm{A, N}} > 5 $\AA\ are being selected by our algorithm despite showing a low combined [SII]-BPT AGN and LIER spaxel fraction f$_{\rm{AL, S}}$, this source is worth to be further investigated. The same is true for the opposite situation. A source with high f$_{\rm{AL, S}}$ coupled with $EW(H\alpha)_{\rm{AL, S}} > 5$\AA\ might not be picked up based on its [NII]-BPT based AGN spaxel fractions f$_{\rm{A, N}}$ because the [SII]-BPT is more sensitive to enhanced emission lines ratios due to shocks which might be related to AGN-driven outflows propagating through the galaxy \citep{Wylezalek_2017}. 

\subsection{Relation to H$\alpha$ Equivalent Width}

In the previous section we show that MaNGA-observed galaxies are generally comprised of two distinct populations in terms of their [NII]-BPT based AGN spaxel fractions f$_{\rm{A, N}}$ and their combined [SII]-BPT AGN and LIER spaxel fractions f$_{\rm{AL, S}}$: one low spaxel-fraction population and a high spaxel fraction population (see black small points in the left panel in Figure \ref{fractions}). We now show that the high spaxel population tends to be correlated with low H$\alpha$ equivalent widths in their [SII]-BPT AGN+LIER spaxels, i.e. $EW(H\alpha)_{\rm{AL, S}} < 5$\AA. This is further visualized in the right panel in Figure \ref{fractions}. This figure shows how f$_{\rm{AL, S}}$ relates to $EW(H\alpha)_{\rm{AL, S}}$ for the whole MaNGA sample (small black data points) and the AGN candidates (large, blue data points). For the whole sample, we again identify the two populations separated in spaxel fraction while the high spaxel fraction population shows $EW(H\alpha)_{\rm{AL, S}} < 5 $\AA. We associate these latter sources with the LIER-type of galaxies \citep{Belfiore_2016}. These sources show large regions of AGN+LIER-type emission as classified by the [SII]-BPT. However, due to the low H$\alpha$ equivalent widths in their [SII]-BPT AGN+LIER spaxels, it is more likely that the bulk of the ionization in these sources is not due to a low-luminosity AGN, but due to other mechanisms, such as ionization through old hot stars \citep[see also][]{Cid-Fernandes_2011}. This is why the bulk of this high spaxel population is not picked up by our AGN selection as the selection deliberately rules out these sources.

\subsection{Relation to H$\alpha$ Surface Brightness}

We now investigate the currently selected AGN candidates with respect to their H$\alpha$ surface brightness. \citet{Zhang_2017} have shown that diffuse ionized gas regions can strongly impact line ratio measurements. The H$\alpha$ surface brightness is a good tracer for diffuse ionized gas regions and due to changes in the hardness of the ionizing spectra (and potentially ionization parameter and metallicity), emission line ratios tend to be enhanced in regions of low H$\alpha$ surface brightness \citep{Haffner_2009, Blanc_2009}. This is often the case in the outskirts of galaxies where the H$\alpha$ surface brightness is lower than $10^{37}$~erg~s$^{-1}$~kpc$^{-2}$. This effects becomes particularly important when working with IFU observations which may map the galaxy out to several effective radii $R_{\rm{eff}}$. 

\begin{figure}
\centering
\includegraphics[scale = 0.47, trim = 0.8cm 0cm 1.55cm 1cm, clip= true]{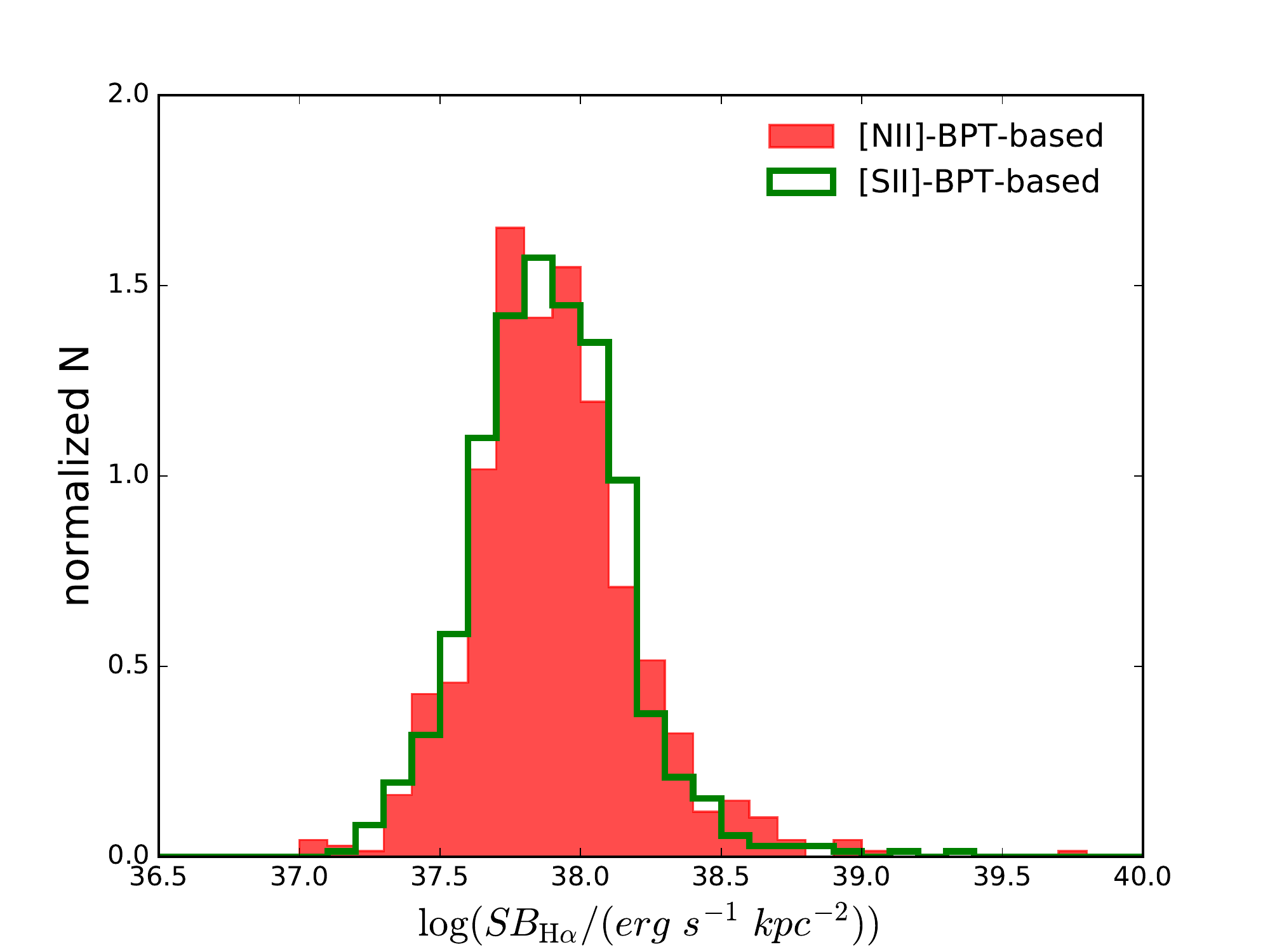}
\caption{Distribution of H$\alpha$ surface brightness: We show the distribution of $\log{(SB(H\alpha)_{\rm{A,N]}}/(erg~s^{-1}~kpc^{-2}))}$ (orange, filled histogram) and $\log{(SB(H\alpha)_{\rm{AL, S}}/(erg~s^{-1}~kpc^{-2}))}$ (green, open histogram). We find a total of 70 sources with either $\log{(SB(H\alpha)_{\rm{A,N]}}/(erg~s^{-1}~kpc^{-2}))}$ or $\log{(SB(H\alpha)_{\rm{AL, S}}/(erg~s^{-1}~kpc^{-2}))} < 37.5$. In regions of such low surface brightness emission line ratios can be enhanced and mimic AGN or LI(N)ER-like line ratios.}
\label{SB}
\end{figure}

We therefore compute both the average H$\alpha$ surface brightness in spaxels that are classified as `AGN' or `Composite' in the [NII]-BPT diagram, SB(H$\alpha$)$_{\rm{A,N}}$, and the average H$\alpha$ surface brightness in spaxels that are classified as `AGN' or `LI(N)ER' in the [SII]-BPT diagram, SB(H$\alpha$)$_{\rm{AL, S}}$. In Figure \ref{SB} we show the distribution of SB(H$\alpha$)$_{\rm{A,N}}$ in red and the distribution of SB(H$\alpha$)$_{\rm{AL, S}}$ in green. We find 70 sources with either either surface brightness measure $< 10^{37}$~erg~s$^{-1}$~kpc$^{-2}$. Visually inspecting these sources shows that indeed the AGN or AGN+LIER selected spaxels trace the outer parts of the galaxy, often in a ring-like pattern. These kind of structures are indicative of diffuse ionized gas indeed impacting the emission line ratio measurements, mimicing AGN and/or LIER-like emission.

\subsection{Significance of BPT classifications}

\begin{figure*}
\centering
\includegraphics[scale = 0.31, trim = 25.5cm 58.65cm 24.cm 7cm, clip= true]{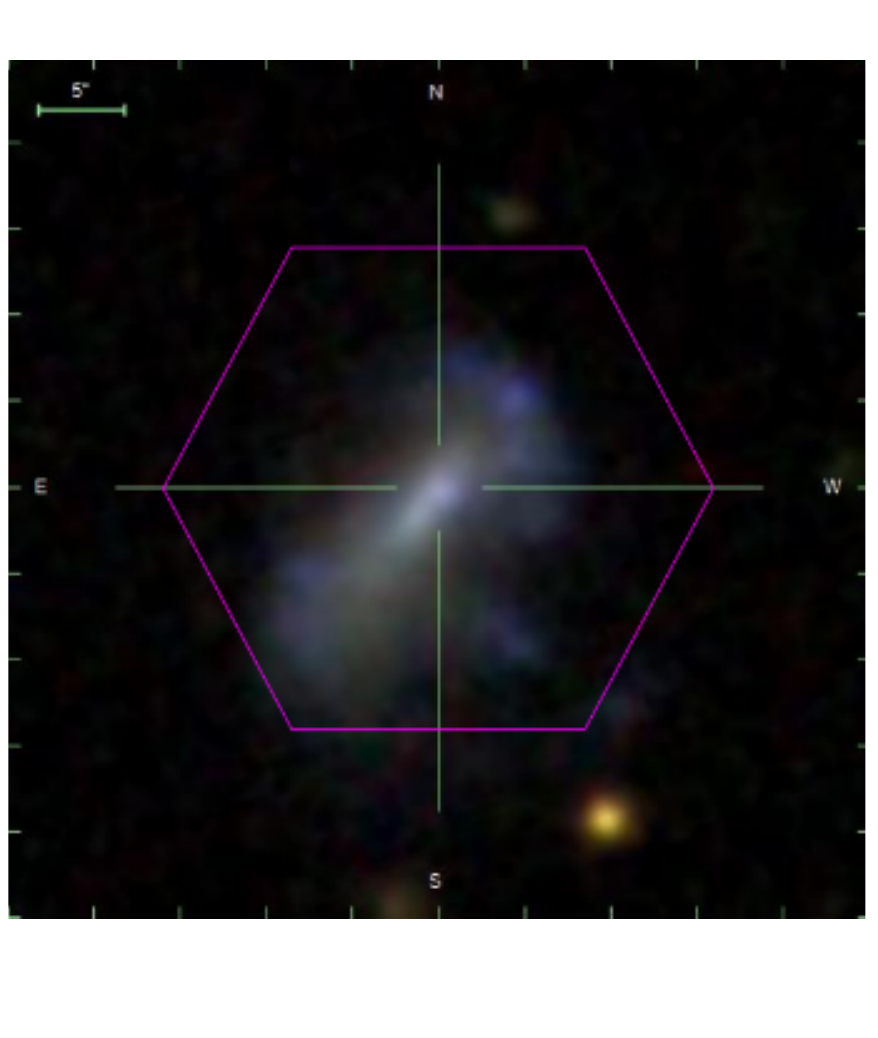}
\includegraphics[scale = 0.31, trim = 6cm 37.7cm 5.6cm 27.5cm, clip= true]{./plots/big_figure/8325_12702.pdf}
\includegraphics[scale = 0.31, trim = 25.5cm 58.65cm 24.cm 7cm, clip= true]{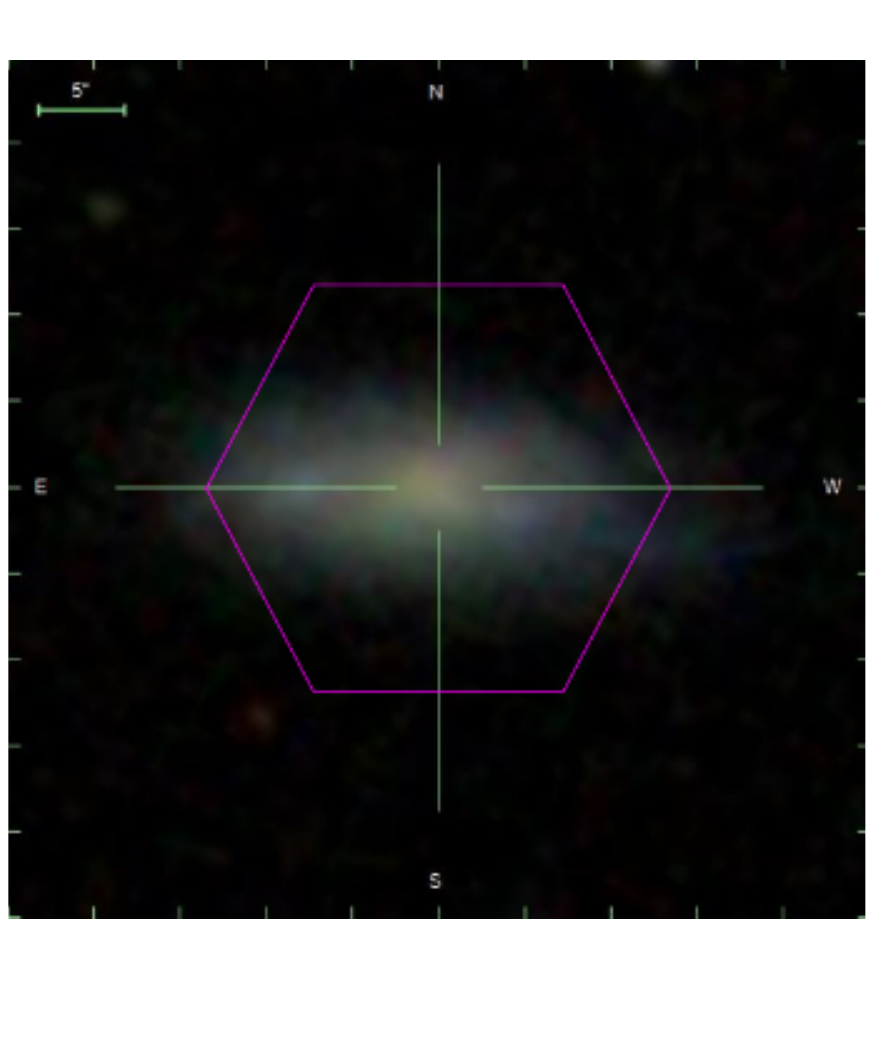}
\includegraphics[scale = 0.31, trim = 6cm 37.7cm 5.6cm 27.5cm, clip= true]{./plots/big_figure/8458_9102.pdf}
\includegraphics[scale = 0.31, trim = 25.5cm 58.65cm 24.cm 7cm, clip= true]{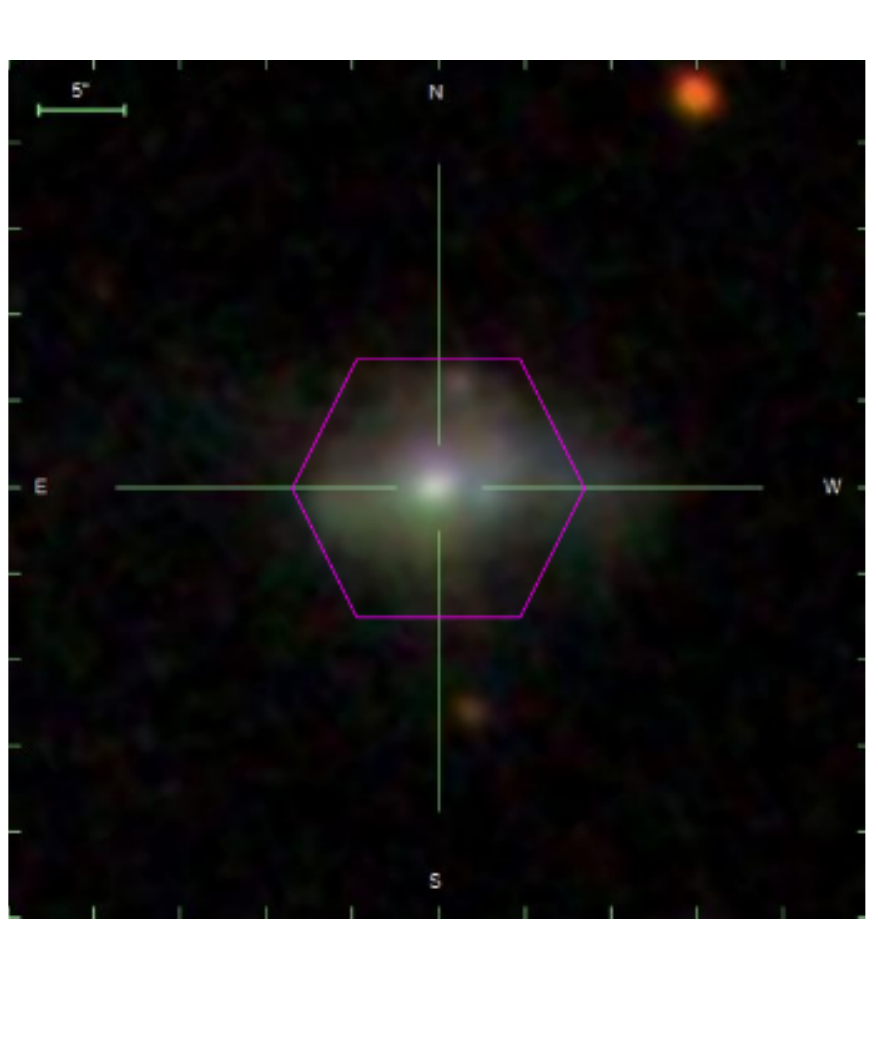}
\includegraphics[scale = 0.31, trim = 6cm 37.7cm 5.6cm 27.5cm, clip= true]{./plots/big_figure/8711_3701.pdf}
\caption{Example BPT maps and plots for three galaxies with high f$_{\rm{AGN+LIER [SII]}}$, but where the spaxels classified as AGN or LI(N)ER in the [SII]-BPT diagram are very close to the demarcation line. The bulk of such galaxies are blue, highly star-forming and of low stellar mass.}
\label{example_low_dbpt}
\end{figure*}

Visually inspecting the selected AGN candidates, we notice a number of sources where the bulk of Seyfert or LI(N)ER classified spaxels are close to the demarcation lines in the BPT maps. We show three examples for such kind of sources in Figure \ref{example_low_dbpt}. As discussed above, the demarcation lines in the BPT diagrams are not hard boundaries. While there are also natural uncertainties in the emission line flux measurements, we note that the average signal-to-noise ratio of the [OIII], H$\beta$, [SII] and H$\alpha$ emission lines in the AGN+LIER dominated spaxels  are 12, 4, 9 and 31, respectively. This shows that these measurements are of high signal-to-noise and the elevated line ratio measurements are not caused by low S/N emission lines. Depending on ISM properties and redshift, emission line ratios may slightly decrease or increase and lead to ambiguous spaxel classifications. This effect is largest close to the demarcation lines. 

\begin{figure}
\centering
\includegraphics[scale = 0.32, trim = 5cm 2cm 5cm 3cm, clip= true]{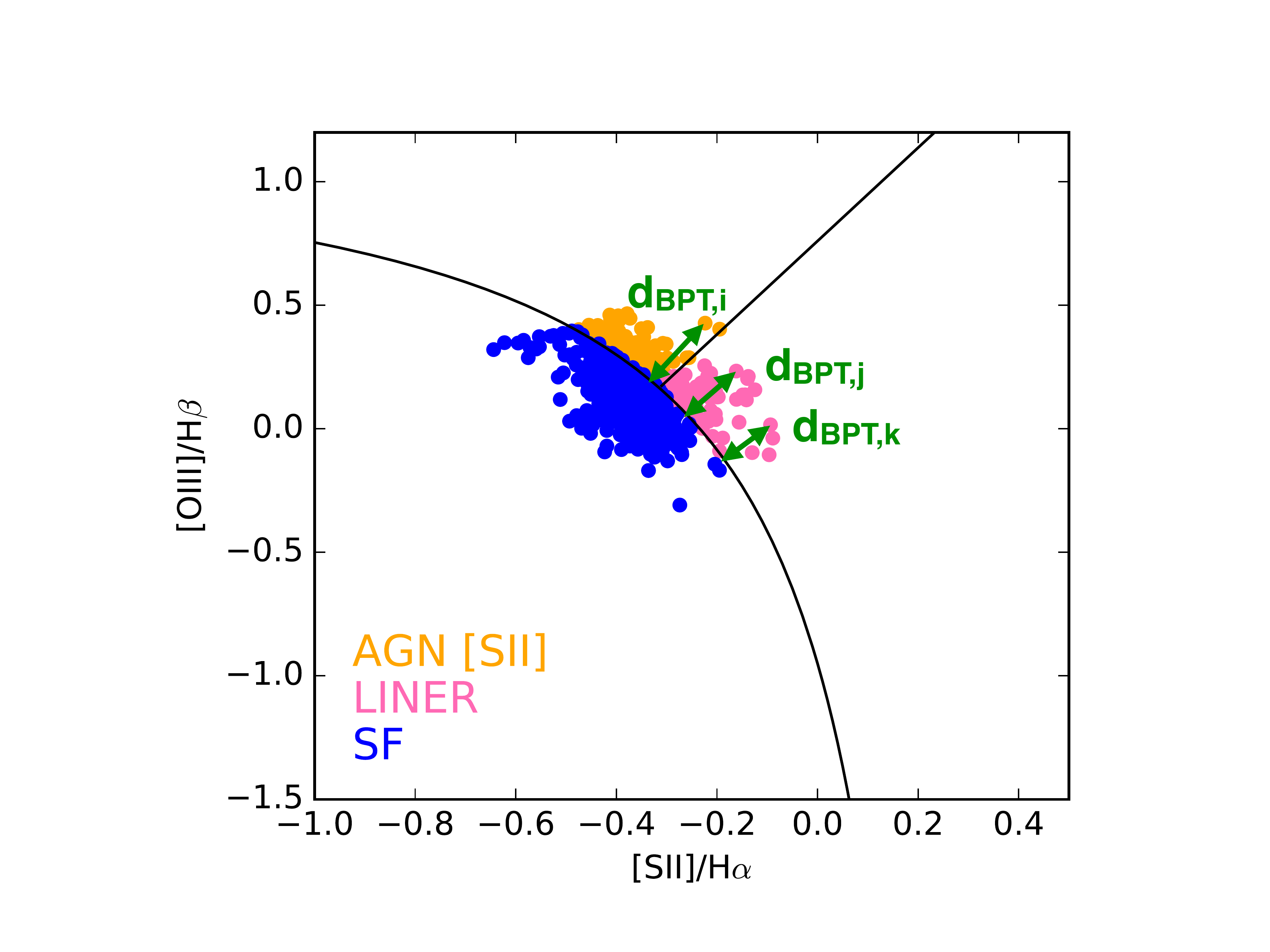} 
\caption{Visualization on how we measure $d_{BPT,i}$ and $d_{BPT}$. For each spaxel in the AGN or LINER region of the [SII] BPT diagram, we measure the distance $d_{BPT,i}$ between the position of the spaxel in BPT space to the star formation demarcation line such that the $d_{BPT,i}$ is minimal. We then compute $d_{BPT}$ by averaging the $d_{BPT,i}$ of the 20\% of the spaxels with the largest $d_{BPT,i}$.}
\label{dbpt_visualization}
\end{figure}

\begin{figure*}
\centering
\includegraphics[scale = 0.49, trim = 0.5cm 0cm 1.7cm 1cm, clip= true]{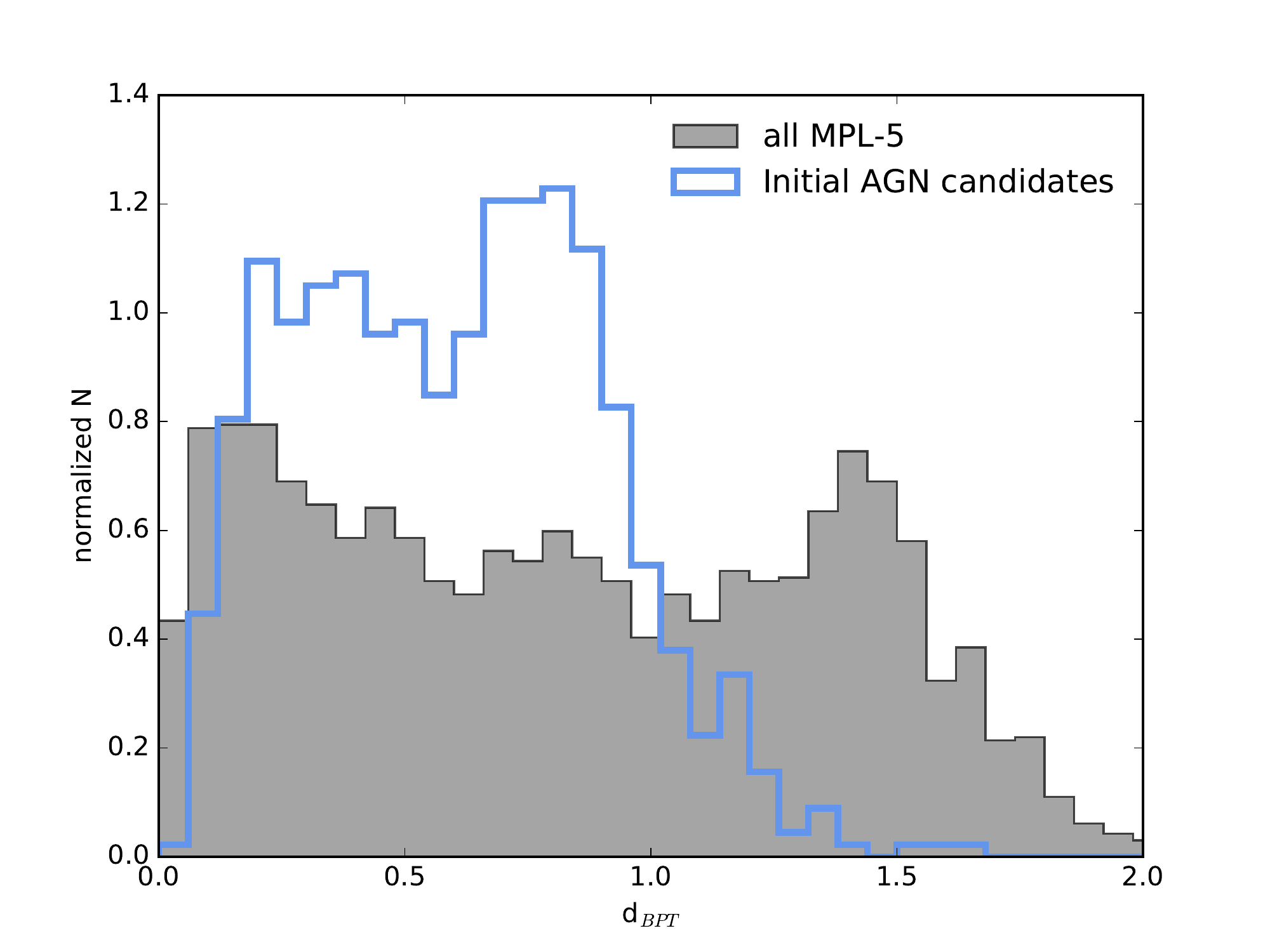} \\
\includegraphics[scale = 0.47, trim = 0.5cm 0cm 1.7cm 1cm, clip= true]{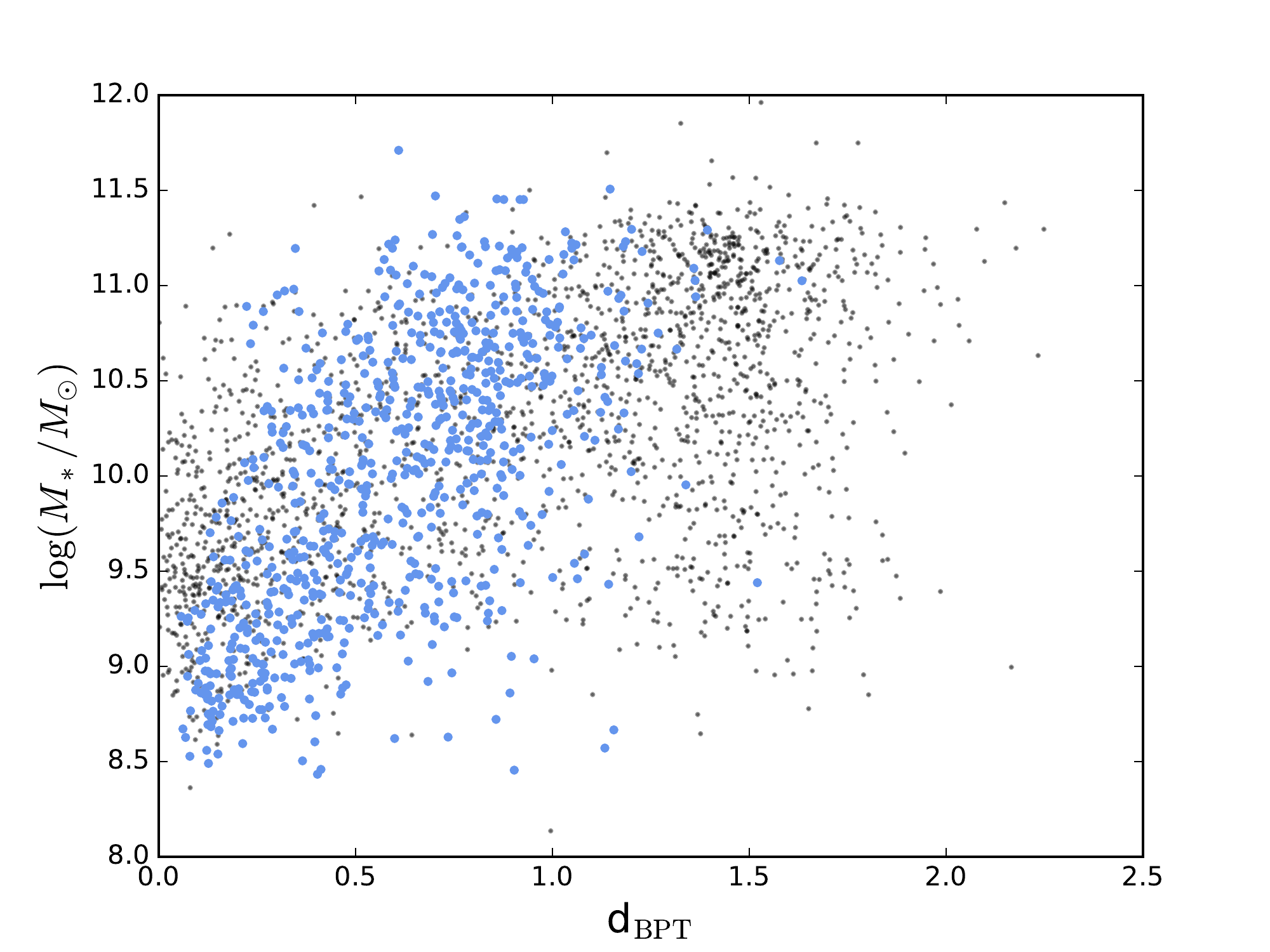}
\includegraphics[scale = 0.47, trim = 0.5cm 0cm 1.7cm 1cm, clip= true]{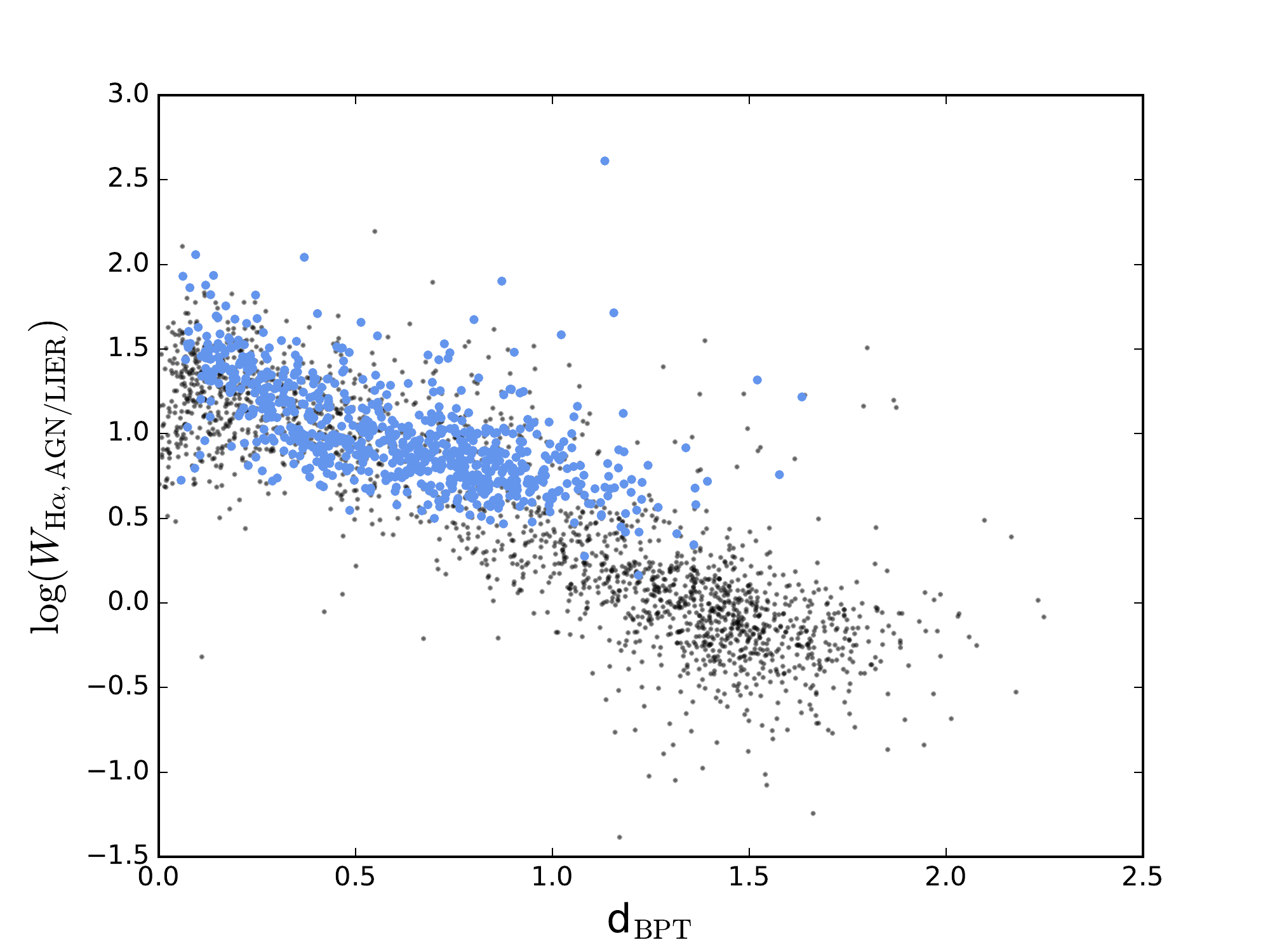}
\caption{\textbf{Upper panel:} Distribution of $d_{BPT}$ for all galaxies in the MaNGA sample (grey, filled histogram) and the initial 746 MaNGA AGN candidates (blue open histogram). $d_{BPT}$ is measured by only regarding 20\% AGN+LI(N)ER spaxels in the [SII]-BPT that are farthest away from the star formation demarcation line. $d_{BPT}$ is then the mean distance of these spaxels from the star formation demarcation line. While the distribution of $d_{BPT}$ for the whole MaNGA sample has two distinct peaks, the distribution of  $d_{BPT}$ for the initially selected AGN excluded the high $d_{BPT}$ which are primarily associated with high significance LIER galaxies that are not AGN. \textbf{Lower left panel:} Stellar mass dependence on $d_{BPT}$, showing that particularly low mass galaxies have small $d_{BPT}$. \textbf{Lower right panel:} H$\alpha$ equivalent width $EW(H\alpha)_{\rm{AL, S}}$ dependence on $d_{BPT}$. Since $EW(H\alpha)_{\rm{AL, S}}$ is also a qualitative measure for the sSFR, this plots shows that the AGN selection is mainly contaminated by low mass galaxies with high sSFR and low $d_{BPT}$.}
\label{dist_line_distribution}
\end{figure*}

To identify sources that through this effect mistakenly made it into our selection, we conduct the following analysis in which we focus on spaxel classifications based on the [SII]-BPT diagrams. In order to measure the significance of a spaxel falling into the Seyfert or LINER regions of the [SII]-BPT diagrams, we measure the `distance' $d_{BPT,i}$ of each Seyfert/LINER-classified spaxel in the BPT diagram from the star formation-demarcation line such that the line connecting the spaxel measurement and the star formation-demarcation line is minimized. In Figure \ref{dbpt_visualization} we visualize how the $d_{BPT,i}$ per spaxel are measured. We then restrict our analysis to the 20\% of the spaxels with the largest distances $d_{BPT,i}$ per galaxy and use the spaxels to compute the mean distance in BPT space $d_{BPT}$ for each source. In Figure \ref{dist_line_distribution} we show the distribution of $d_{BPT}$ for all galaxies in the MaNGA sample (grey filled histogram) and the AGN candidates (blue unfilled histogram). The distribution of all galaxies in the MaNGA sample has two prominent peaks at low $d_{BPT} \sim 0.15$ and at high $d_{BPT} \sim 1.5$. The high $d_{BPT}$ population, where the majority of the spaxels lie well in the Seyfert/LINER region of the [SII]-BPT is related to the high spaxel fraction/low H$\alpha$ equivalent width population of galaxies discussed above and represents the bona-fide LI(N)ER galaxies that are most likely not associated with AGN. 

The distribution for $d_{BPT}$ of the AGN candidates is relatively flat over the range $0.1 < d_{BPT} < 1$ with a tail towards high $d_{BPT}$. This distribution shows that the BPT spaxel classifications of the AGN candidates is quite significant in that only a small fraction of AGN candidates shows low $d_{BPT}$. There are 126 sources within our AGN-selected sample with $d_{BPT} < 0.3$. In these galaxies the bulk of the AGN and/or LI(N)ER classified spaxels lie very close to the star formation-demarcation line. Such distances from the demarcation line can be reached by changes in the ionization parameter or metallicity. Although we have made efforts to exclude galaxies that are dominated by such effects by implying additional cuts on the H$\alpha$ equivalent width and surface brightness, these sources may have been missed. Visually inspecting the AGN candidates with $d_{BPT} < 0.3$, we notice that in the majority of the sources, AGN and LI(N)ER spaxels indeed seem to be the high emission line ratio tails of the star forming spaxel distribution in the BPT diagrams. 

We also notice that the many of the sources are small, blue (in term of their SDSS $gri$ composite image) with strong nebular emission lines (Figure \ref{example_low_dbpt}).  In Figure \ref{dist_line_distribution} we therefore also show that $d_{BPT}$ is a strong function of both stellar mass and the H$\alpha$ equivalent width $EW(H\alpha)_{\rm{AL, S}}$. Sources with $d_{BPT} < 0.3$ are primarily low-mass (median mass $\log{(M_*/M_{\sun})} = 9.5$) with high H$\alpha$ equivalent widths $EW(H\alpha)_{\rm{AL, S}} \sim 25$\AA.

The fact that $d_{BPT}$ is both a strong function of stellar mass and H$\alpha$ equivalent width $EW(H\alpha)_{\rm{AL, S}}$ for both all MaNGA galaxies and the AGN candidates within MaNGA is not necessarily a surprise. As can be seen in Figure \ref{dist_line_distribution}, the AGN candidates are primarily sources with $d_{BPT} < 1.5$, whereas the bulk of the non-AGN candidates have $d_{BPT} > 1.5$. This latter category are the same, low H$\alpha$ equivalent width, high spaxel fraction sources discussed above that we associate with typical LIER-type galaxies, in which the observed emission line ratios are primarily a result of photoionization through hot, evolved stars. This also means that these galaxies are dominated by old stellar populations and have already built up most of their stellar mass. This is why these high $d_{BPT}$, low $EW(H\alpha)_{\rm{AL, S}}$ galaxies make up a large fraction of the high mass galaxies. In galaxies that are not dominated by AGN, the H$\alpha$ equivalent width is also a quantitative measure of the specific star formation rate (sSFR). This follows from the fact that the H$\alpha$ equivalent width is the ratio of a star-formation indicator (H$\alpha$ line flux) and the stellar continuum at $\lambda_{rest} \sim 6563$\AA\, a reasonable measure of stellar mass \citep{Marmol_Queralto_2016}. The higher mass a galaxy has, the clearer the signatures of the old stellar population becomes, i.e. the more significant the LIER signatures and the higher $d_{BPT}$. 

In contrast, low $d_{BPT}$ values arise in low mass galaxies with high H$\alpha$ equivalent widths $EW(H\alpha)_{\rm{AL, S}}$, i.e. galaxies with high sSFR (if no AGN is present). In many of these galaxies, the AGN and LI(N)ER signatures are primarily detected at the edges of the MaNGA-recovered footprint of the galaxy, where shocks from starburst-driven winds may contribute in elevating the line ratios in regions where also contributions from diffuse ionized gas are relevant, despite our efforts to minimize these effects. Based on a stacking analysis of 49 edge-on, late-type galaxies observed within MaNGA, \citet{Jones_2017} have investigated extended extra-planar diffuse ionized gas out to several $R_{eff}$ above the midplane. Similar to \citet{Zhang_2017}, they show that both [NII]/H$\alpha$ and [SII]/H$\alpha$ can be higher at large distances $R > 2.5R_{eff}$. The effect is more significant for [SII]/H$\alpha$ compared to [NII]/H$\alpha$. 

On the other hand, using optical spectra from the SDSS Data Release 6, \citet{Brinchmann_2008} has shown that galaxies with significant contributions of Wolf-Rayet features to their spectra, can show elevated BPT emission line ratios and populate the AGN region in the [NII] and [SII] BPT diagram. Wolf-Rayet stars are hot, massive stars whose spectra are dominated by broad, strong emission lines associated with massive circumstellar shells expanding outwards. They can provide valuable information about recent star formation activity in galaxies since they typically start to appear about $2\times10^6$~years after a star formation episode and disappear within $5\times10^6$~years \citep{Brinchmann_2008}. The first early reports of the detection of Wolf-Rayet stellar features in galaxy spectra were of starbursting low-metallicity galaxies, typically blue compact dwarfs \citep{Kunth_1983}, similar to the galaxies with low $d_{BPT}$ values in this work. The systematic search for Wolf-Rayet features in galaxies by \citet{Brinchmann_2008} confirms this early impression: most of the galaxies with significant Wolf-Rayet features show strong star formation activity with high sSFR. Additionally, the fraction of galaxies showing Wolf-Rayet features is highest in low-metallicity, blue (in terms of their $g-r$ color) galaxies. The authors relate the elevated line ratios observed in some Wolf-Rayet galaxies with a higher effective ionisation parameter. While \citet{Brinchmann_2008} show that only a small fraction of the galaxies with Wolf-Rayet features contaminate the AGN-part of the [SII] and [NII] BPT diagram, we remind the reader that this work had been based on SDSS single fibre measurements. A full census of Wolf-Rayet signatures in MaNGA galaxies is beyond the scope of this paper. But contamination of emission-line ratio measurements due Wolf-Rayet signatures appears to be more significant in galaxy IFU surveys compared to single fibre surveys. 

This analysis leads us to the conclusion that the bulk of our initial AGN candidates with $d_{BPT} < 0.3$ are unlikely to host true AGN. In addition to the `known' contaminants in AGN classification (contamination from diffuse ionized gas regions and contamination from LIER emission related to old, hot stars), we conclude that contamination from hot, young stars (such as Wolf-Rayet stars) contributes as well, and therefore has to be addressed in IFU-based AGN selection algorithms.

\section{Analysis of MaNGA AGN Candidates}

\subsection{Refined AGN Selection}

Based on the analysis of our initial AGN selection above, we conclude that when selecting AGN candidates in large IFU surveys such as MaNGA, additional constraints have to be taken into account that are mostly due to the resolved nature of the observations and radial gradients in emission line ratios that do not allow for simple adaptation of BPT diagnostics.

Based on the analysis above, we therefore propose the following refined AGN selection criteria, which have been optimized for the MaNGA dataset:

\begin{itemize}[align=left,leftmargin=*,labelsep=1ex]
\item f$_{\rm{A, N}} > 10$ \% and $EW(H\alpha)_{\rm{A, N}} > 5 \AA$

\vspace{0.2cm}
AND
\vspace{0.1cm}

\item f$_{\rm{AL, S}} >15$\% and $EW(H\alpha)_{\rm{AL, S}} > 5 \AA$

\vspace{0.2cm}
AND
\vspace{0.1cm}

\item  $\log$(SB(H$\alpha$)$_{\rm{A,N}}$/(erg~s$^{-1}$~kpc$^{-2}$)) $>37.5$ OR \newline $\log$(SB(H$\alpha$)$_{\rm{AL,S}}$/(erg~s$^{-1}$~kpc$^{-2}$)) $>37.5$

\vspace{0.2cm}
AND
\vspace{0.1cm}

\item $d_{\rm{BPT}} > 0.3$

\end{itemize}

\begin{figure*}
\centering
\includegraphics[scale = 0.31, trim = 25.5cm 58.65cm 24.cm 7cm, clip= true]{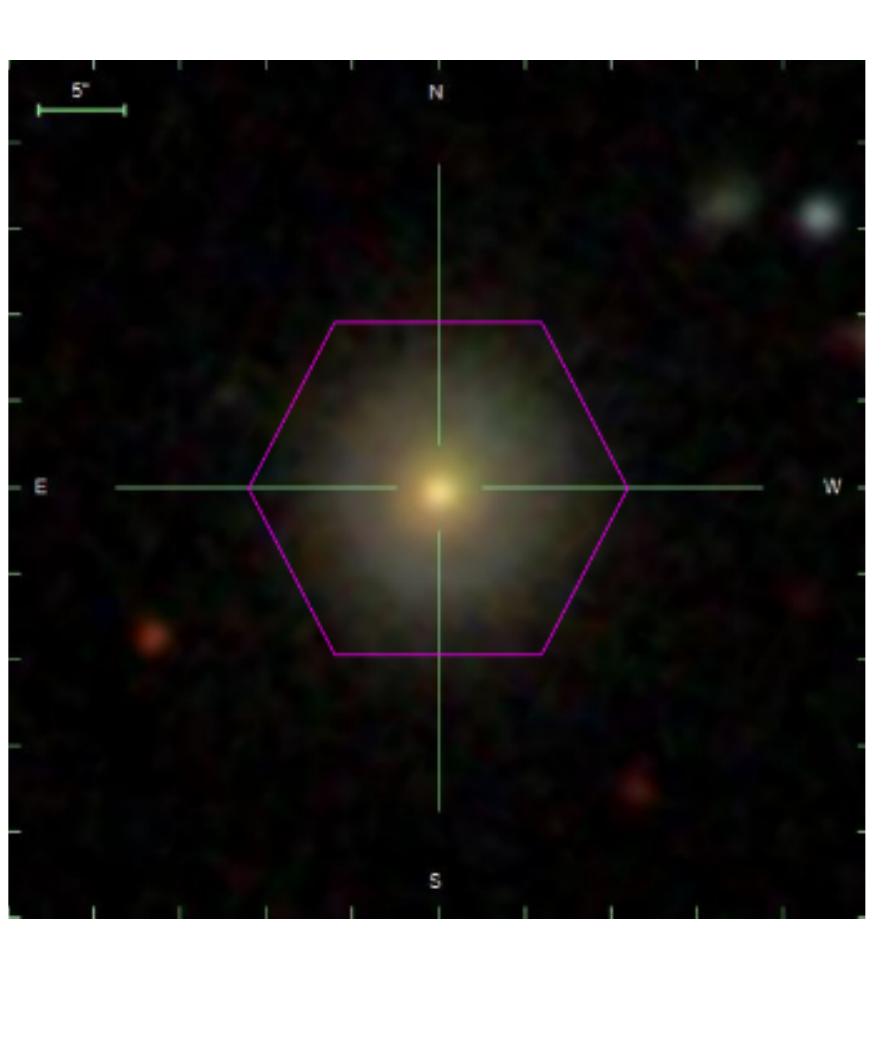} 
\includegraphics[scale = 0.31, trim = 6cm 37.7cm 5.6cm 27.5cm, clip= true]{./plots/big_figure/8978_6104.pdf}
\includegraphics[scale = 0.31, trim = 25.5cm 58.65cm 24.cm 7cm, clip= true]{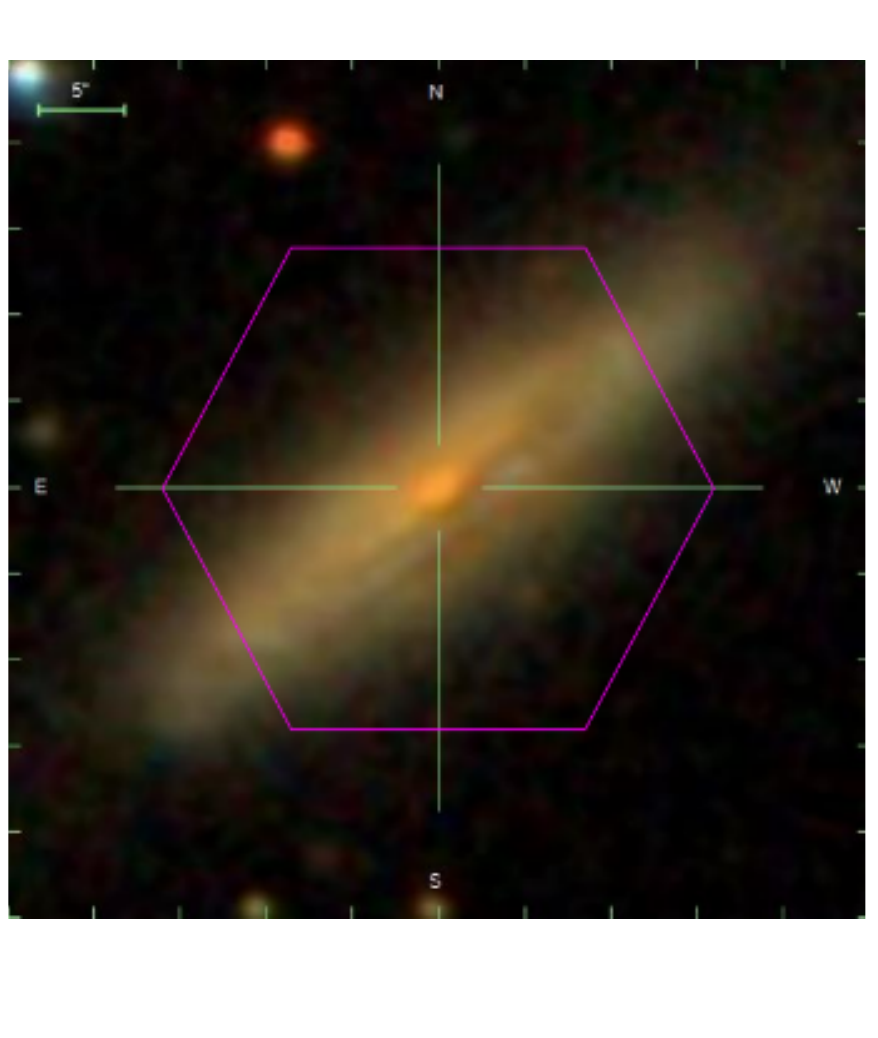}
\includegraphics[scale = 0.31, trim = 6cm 37.7cm 5.6cm 27.5cm, clip= true]{./plots/big_figure/7977_12704.pdf}
\includegraphics[scale = 0.31, trim = 25.5cm 58.65cm 24.cm 7cm, clip= true]{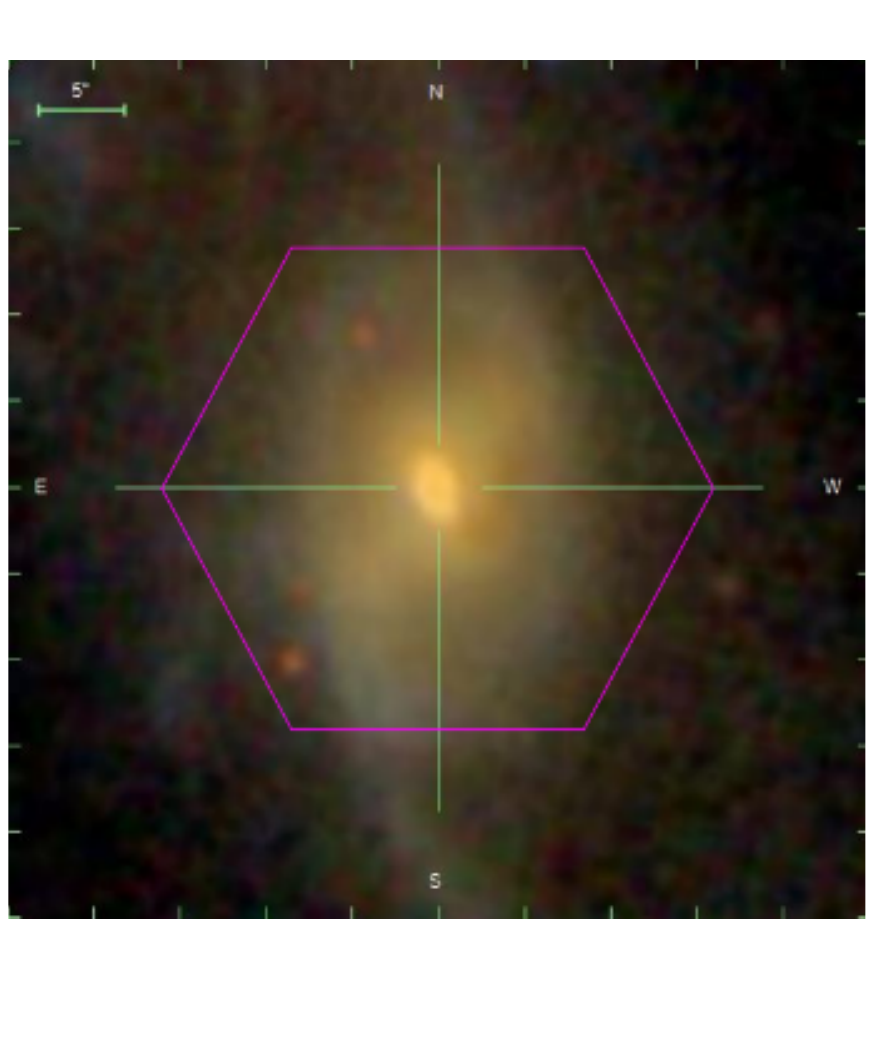}
\includegraphics[scale = 0.31, trim = 6cm 37.7cm 5.6cm 27.5cm, clip= true]{./plots/big_figure/7977_12705.pdf}
\caption{Example BPT maps and plots for three galaxies in our final AGN selection. In addition to the SDSS composite $gri$ image, we show the MaNGA-based resolved [NII] and [SII] BPT maps and the corresponding BPT diagrams for each spaxel. The green circle illustrates the size of the $3\arcsec$ fibre that was used to obtain a spectrum of the galaxy in SDSS I-III. While the galaxy in the upper row had been classified as an AGN based on the single fibre observations prior to MaNGA, the galaxies in the middle and lower row had not been selected as AGN candidates based on the single-fibre spectra.}
\label{AGN_selection_final}
\end{figure*}

These criteria select 303 unique AGN candidates. This corresponds to an AGN fraction of $\sim 10\%$. We note that the average signal-to-noise ratio in the individual lines per spaxel are $\sim 7, 8, 8, 12$ and 42 for the H$\beta$, [OIII], [NII], [SII] and H$\alpha$ lines, respectively. We will refer to this sample as the `final AGN candidates'. We note that the AGN+LINER classified spaxels of the `final' AGN candidates are generally found within 1~R$_{eff}$. That shows that this refined AGN selection cleans the sample of sources where contaminating effects beyond 1~R$_{eff}$ may lead to a wrong BPT classification. We show example BPT maps and plots for three of our final AGN candidates in Figure \ref{AGN_selection_final}.

\subsection{Observed Galaxy Radii}

A major part of both the initial and the refined AGN selection criteria are based on measuring AGN spaxel fractions. We apply this threshold that is only dependent on observed size to \textit{(i)} minimize the contamination from galaxies that would pass our AGN selection criterion due to measurements based on only a few single spaxels (which might represent a large physical region) and \textit{(ii)} allow for an easy application to the whole MaNGA sample in subsequent years.
But due to the design of the fibre bundles, the required spaxel threshold of 15\% in our initial AGN selection criterium and 10\%/15\% in our refined AGN selection corresponds to different spatial fractions depending on whether the source belonged to MaNGA primary sample (fibre bundle radius corresponds to $\sim $1.5~R$_e$) or the secondary sample (fibre bundle radius corresponds to $\sim $2.5~R$_e$). We therefore investigate to what extent this simple fraction threshold biases our sample selection.

\begin{figure}
\centering
\includegraphics[scale = 0.47, trim = 0.8cm 0cm 1.7cm 1cm, clip= true]{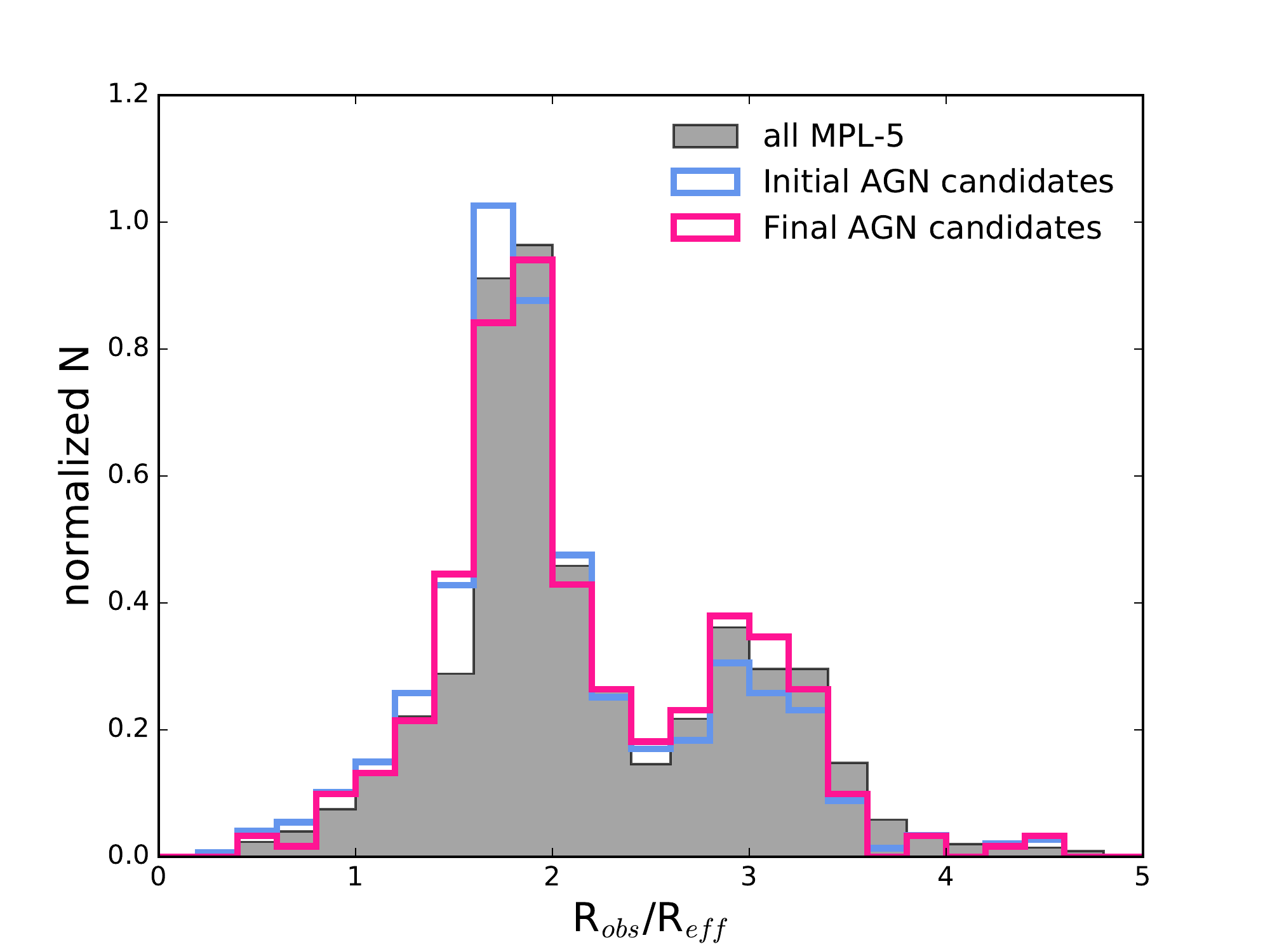}
\caption{Normalized distribution of observed galaxy radii normalized by their R$_{eff}$ for all galaxies in the MaNGA sample (filled, grey histogram), initial AGN candidates (blue) and final AGN candidates (pink). While our initial AGN selection was biased towards galaxies with low $R_{obs}/R_{eff}$, the final AGN selection overcomes this bias. A two-sided KS test between the all MaNGA distribution and the final AGN distribution suggests that the two distributions are drawn from the same underlying distribution.}
\label{reff_distribution}
\end{figure}

Figure \ref{reff_distribution} shows the normalized distributions of observed galaxy radii $R_{obs}$, i.e. the fibre bundle radius, to the effective radius of the observed galaxy $R_{eff}$ for all galaxies in the MaNGA sample and the 746 initial AGN candidates and the 303 final AGN candidates. While the distributions between the whole MaNGA sample and the initial AGN sample (blue) differed significantly ($p$-value = 0.001 based on a two-sided Kolmogorov-Smirnoff test), the distributions between the final AGN sample (pink) and the whole MaNGA sample are likely drawn from the same distribution ($p-$value = 0.83). This shows that our initial AGN selection the percentage-based spaxel threshold biased the selection towards smaller $R_{obs}/R_{eff}$ and therefore to lower redshift sources for which a smaller areal fraction of the galaxy is mapped on average due to larger angular sizes of the galaxies. That led to the selection of many AGN candidates with only marginal signatures. The refined AGN selection criteria manage to circumvent this caveat and re-produce the overall MaNGA distribution well.

\subsection{Stellar Mass Distribution}

\begin{figure}
\centering
\includegraphics[scale = 0.47, trim = 0.8cm 0cm 1.5cm 1cm, clip= true]{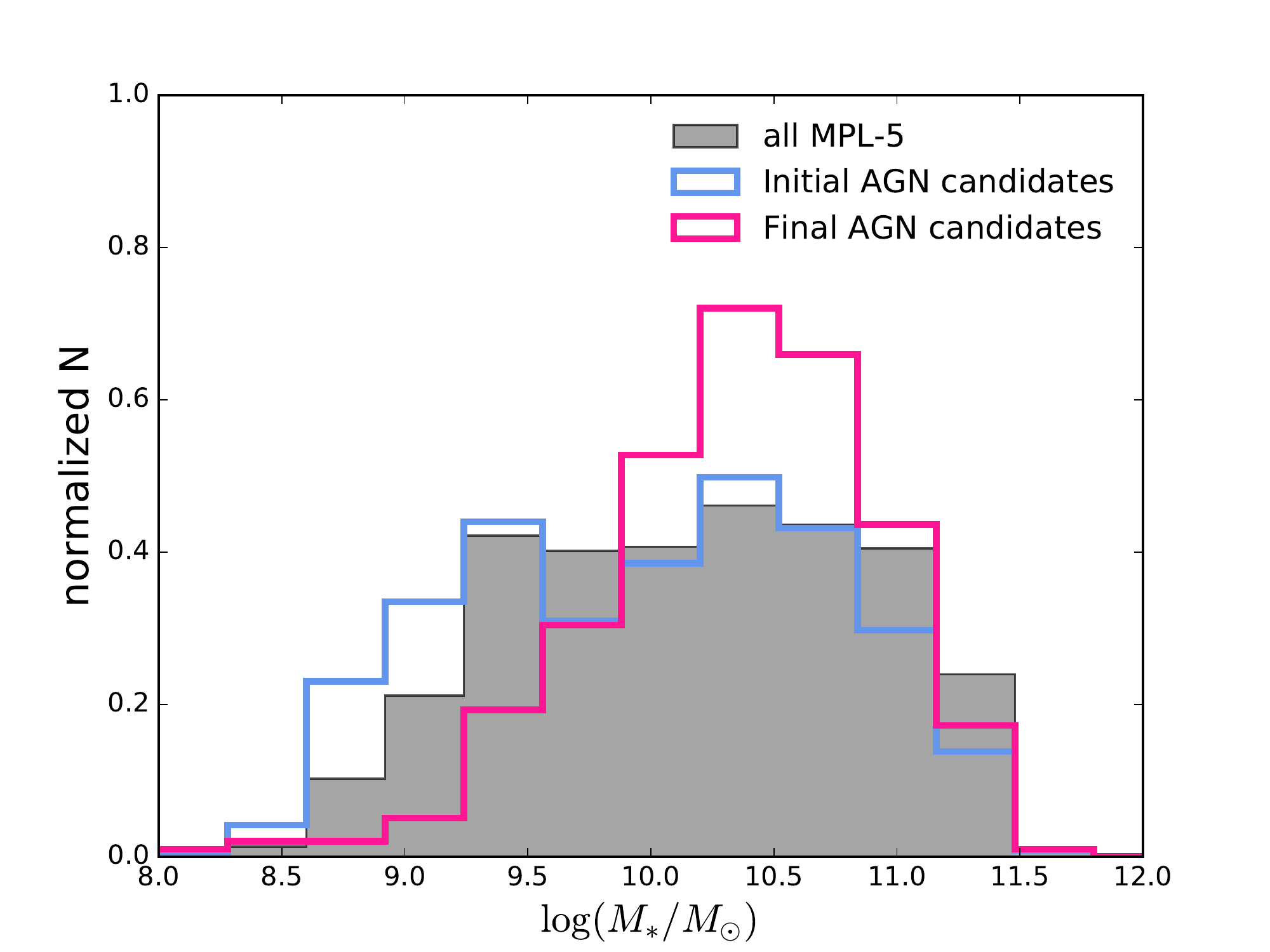}
\caption{Normalized distribution of the stellar mass for all galaxies in MaNGA (filled, grey histogram), initial AGN candidates (blue) and final AGN candidates (pink). While our initial AGN selection was biased towards galaxies with low stellar masses, the final AGN selection peaks at $\log(M_{*}/M_{\sun}) \sim 10.4$.}
\label{mass_distribution}
\end{figure}

By design of the MaNGA sample, the distribution of stellar masses is relatively flat between $9 < \log(M_{*}/M_{\sun}) < 11.5$ (Figure \ref{mass_distribution}). The initial AGN selection was biased towards lower mass galaxies with high specific star formation rates in which emission line ratios seem to be enhanced and mimic AGN or LIER-like emission. After accounting for these wrongly classified AGN and adopting our refined AGN selection criteria, the stellar mass distribution of our AGN candidates peaks at $\log(M_{*}/M_{\sun}) \sim 10.4$ (see pink histogram in Figure \ref{mass_distribution}). This is 0.2~dex higher than the mean stellar mass of the whole MaNGA sample but does not include the most massive galaxies within MaNGA. As we have shown in previous sections, the high mass galaxies within MaNGA are mainly dominated by LI(N)ER-like emission that is related to low H$\alpha$ equivalent widths. This kind of emission is in most cases not related to ionization through an AGN. That said, our minimal required equivalent width naturally biases our selection agains massive, gas poor galaxies with weak or no optical emission lines even if they contain an AGN. Unless the central nucleus is seen directly as a type 1 AGN, optical AGN identification relies completely on gas illuminated by the hidden nucleus and therefore preferentially selects gas-rich objects, unless they are completely enshrouded. We discuss this bias against high mass, gas poor galaxies in more detail in Section 5.

\section{Comparison to previous AGN Classifications}

\subsection{Classification based on SDSS single-fibre spectra}

We now investigate how the selection of our 303 AGN candidates relates to selection that would have been made based on SDSS-III single-fibre observations alone. To that purpose we utilize the emission-line flux measurement catalog from the Portsmouth Group \citep{Thomas_2013} which provides emission-line measurements for SDSS galaxies observed by the DR12 release date. We first cross-match all galaxies observed within MaNGA with the Portsmouth catalogs and obtain 2463 matches for the 2727 galaxies in the MaNGA sample. This might be surprising since MaNGA targets are selected from the SDSS DR7 MAIN galaxy sample \citep[for details see][]{Bundy_2015} which means that single-fibre spectra exist for all MaNGA targets. The reason why not all MaNGA galaxies have a match in the Portsmouth emission line database is due to various quality cuts that have been imposed to the Portsmouth products. We therefore restrict the following analysis to galaxies with single-fibre emission line flux measurements in the Portsmouth catalogs.

\begin{figure*}
\centering
\includegraphics[scale = 0.4]{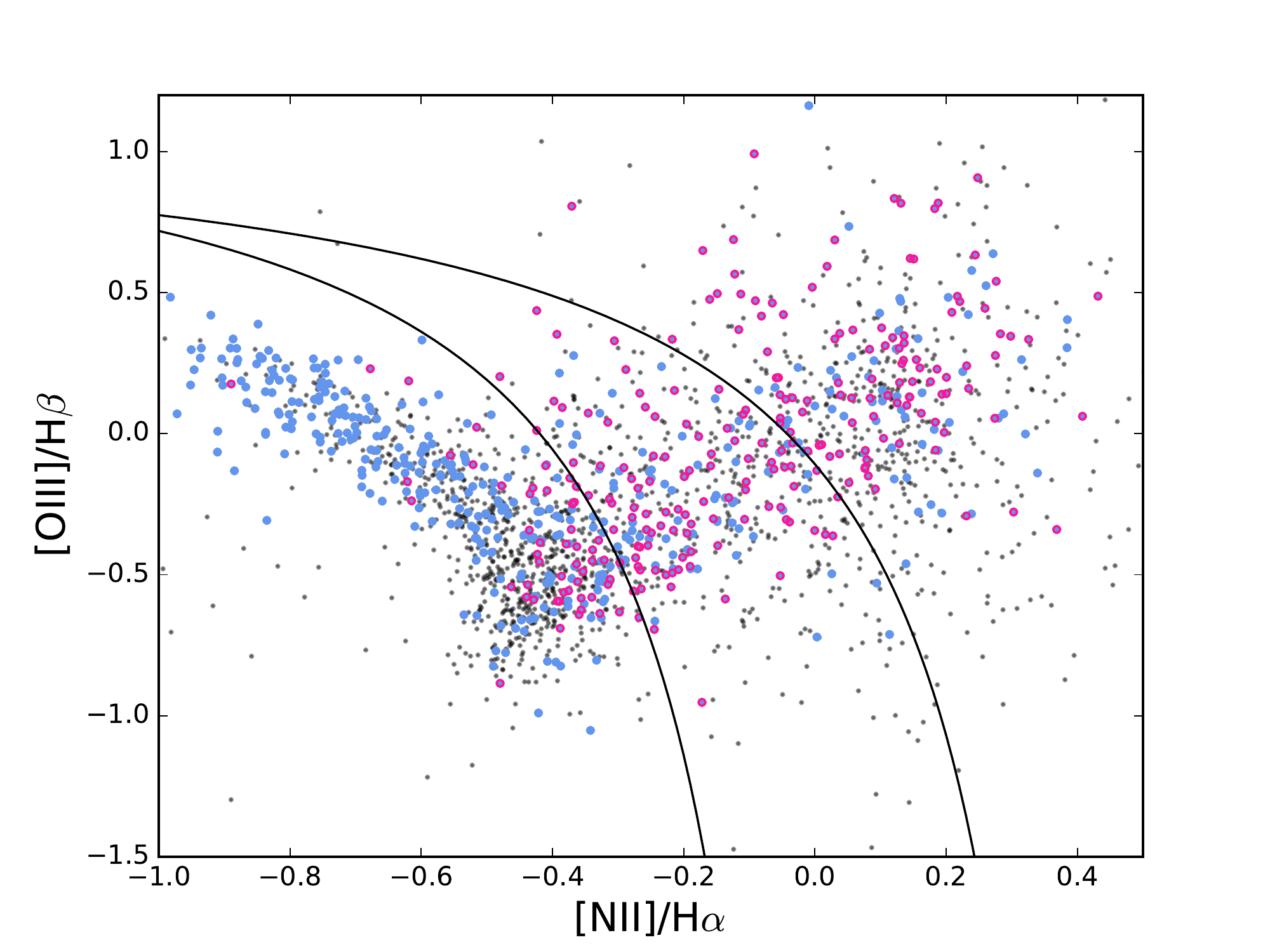}
\includegraphics[scale = 0.4]{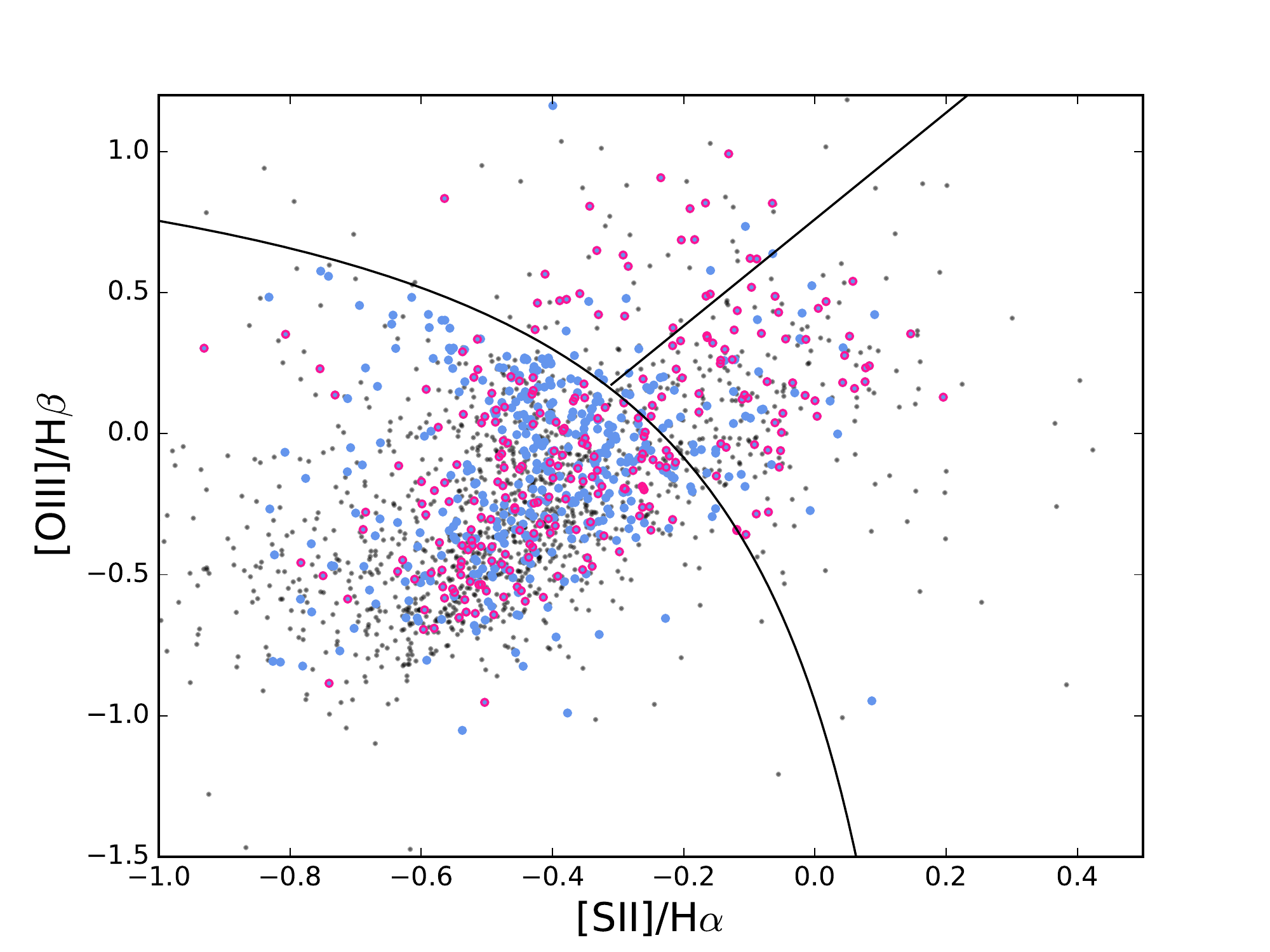}
\caption{\textbf{Left:} [NII]-BPT based on SDSS-III single fibre measurements for galaxies observed within MaNGA (grey datapoints), the initial AGN candidates selected in this paper (blue) and the final AGN candidates (pink). The final AGN selection decreases the number of contaminants from highly star forming galaxies. \textbf{Right:} [SII]-BPT based on SDSS-III single fibre measurements for galaxies observed within MaNGA (grey datapoints), the initial AGN candidates selected in this paper (blue) and the final AGN candidates (pink). These distributions show that many AGN may have previously been missed in single-fibre/single-slit observations and that large IFU surveys offer a new window in finding hidden, offset or changing look AGN.}
\label{BPT_SF}
\end{figure*}

Figure \ref{BPT_SF} shows the distribution of MaNGA galaxies in the [NII] and [SII] BPT diagrams given the emission-line fluxes from single-fibre observations. The small black data points show the positions for all MaNGA-Portsmouth matched galaxies while the larger blue data points show the position of the initial MaNGA AGN candidates and the large pink data points show the MaNGA-selected AGN based on our refined selection criteria. The distribution of sources in the BPT plots based on the single-fibre measurements recover the well-known SF/AGN and SF/AGN/LINER branches in the [OIII]/H$\alpha$ vs. [NII]/H$\beta$ and  [OIII]/H$\alpha$ vs. [SII]/H$\beta$ space, respectively. Based on the single-fibre measurements, 588 galaxies would have been classified as AGN, 459 as Composite and 1065 as star-forming according to the [NII]-BPT, while based on the [SII]-BPT measurements, 86 would have been classifies as AGN, 317 as LINER and 1570 as star-forming galaxies. For the remaining galaxies, emission line ratios could not be computed since at least one emission line flux was reported to be zero in the Portsmouth catalog. 

The MaNGA AGN candidates span a wide range in single-fibre classifications. In our sample of 303 final MaNGA AGN candidates, there are 173 galaxies that would not have been selected as AGN candidates based on the single-fibre measurements and the [SII] BPT diagram, but do show convincing AGN signatures in the MaNGA IFU maps. We show examples for two of such sources in the lower two rows of Figure \ref{AGN_selection_final}. In addition to the resolved MaNGA BPT maps, we also show the size of the $3 \arcsec$ optical fibre showing the area of the galaxy based on which galaxies in SDSS have previously been classified. In both galaxies the central $3\arcsec$ do not show clear AGN-like line ratios, potentially due to heavy obscuration along the line of sight altering emission line ratios or `hiding' AGN photoionized regions.

In the source in the middle row, the AGN-like signatures on larger scales have a cone-like morphology, similar to the Cone Source in \citet{Wylezalek_2017}. Above and below the plane of the disk, where dust obscuration is lower, the AGN ionization cones become apparent. In the source in the bottom row, the central $3\arcsec$ are completely dominated by SF-like emission, potentially related to nuclear starbursts close to the galactic center. But again, beyond the central $3\arcsec$, AGN-photoionized gas dominates the spectral signatures in the galaxy, revealing a `hidden' AGN in this source. Alternatively, the AGN in this galaxy might have recently turned off \citep{Shapovalova_2010, McElroy_2016, Ichikawa_2017} such that gas in the inner parts of the galaxy is not photo-ionized by the AGN anymore but the spectral signatures are again dominated by the star formation processes in the center of the galaxy \citep{Keel_2012, Keel_2015}. The AGN-like signatures that are now apparent in the MaNGA maps then represent relic AGN-photoionized regions or AGN light echoes. In Wylezalek et al. in prep. and Flores et al. in prep., we investigate true nature of such AGN candidates using both multi-wavelength approach and an assessment of the gas kinematics in these sources.  

On the other hand, there are 324 of sources which had been classified as AGN or LI(N)ER based on the single-fibre measurements, but are not in our sample of MaNGA AGN candidates. The reason for this is twofold. Looking at the [SII]-BPT plot, there are 261 LI(N)ER sources that we do not select as AGN candidates. We reject most of these sources based on their low H$\alpha$ equivalent widths. As mentioned earlier, several distinct classes of objects overlap in the LI(N)ER region of the [SII] BPT diagram, and imposing a threshold on the H$\alpha$ equivalent width allows to distinguish between true LINER galaxies and `LIERs' \citep{Cid-Fernandes_2011, Belfiore_2016}. However, there are also 63 sources in the AGN-region of the [SII]-BPT that we do not select as AGN candidates based on the refined MaNGA AGN selection. Most of these sources have also been rejected based on their low H$\alpha$ equivalent width. In some of these sources the low H$\alpha$ equivalent width may point to a different ionization mechanism other than through AGN. But a low H$\alpha$ equivalent width is also expected for gas-poor, massive galaxies in which there is just not enough gas to be ionized. Indeed, the mean stellar mass of the `missed' AGN galaxies in the [SII]-BPT is $\log(M_{*}/M_{\sun}) = 10.6$, 0.2~dex higher than the mean stellar mass of the MaNGA-selected AGN candidates. The advantage of single-fibre observations in such cases is that emission lines may be observed with a higher signal-to-noise ratio because all flux in the central $3\arcsec$ is summed. 

We note that part of the discrepancy between the single-fibre classifications and the MaNGA-based classifications may be due to differences in the stellar continuum subtraction done by the Portsmouth Group \citep{Thomas_2013} from that done by the MaNGA Data Analysis Pipeline. We therefore repeat the above described analysis using MaNGA measurements alone. We first measure the relevant emission line fluxes in 3~arcsec apertures (the size of the SDSS-III fibres) in the MaNGA maps. These MaNGA-based 3 arcsec emission line fluxes generally agree well with the emission line fluxes based on the Portsmouth catalog resulting in similar distributions of flux ratio measurements in in the [NII]- and [SII]-BPT diagrams. This shows that the findings described above are not driven by any systematic differences between the single-fibre measurements and the MaNGA-based 3~arcsec aperture measurements. 

\subsection{Type-1 AGN}

The MaNGA Data Analysis Pipeline has been designed to fit and measure galaxy stellar continua, emission and absorption lines but has not fully been optimized to deal with extreme cases such as reliably identifying and measuring the broad emission lines in type-1 AGN. To asses how well the MaNGA Data Analysis Pipeline and therefore our selection algorithm recovers type-1 AGN, we cross-correlate all MaNGA observed galaxies with the type-1 AGN catalog from \citet{Oh_2015}. \citet{Oh_2015} developed a robust algorithm to select type-1 AGN at $z < 0.3$ from SDSS DR7. In addition to the type-1 AGN that were already flagged by the SDSS pipeline, the authors identified new type-1 AGN based on improved emission line measurements tailored to account for weak broad emission lines. Therefore, the newly identified type-1 AGN in \citet{Oh_2015} are predominantly of lower luminosity than the previously known type-1 AGN. 

We find 19 matches between the catalog from \citet{Oh_2015} and the MaNGA catalog of which 11 sources are in our final selection of AGN candidates. The 8 type-1 AGN missed in our selection were either missed due to bad emission line flux measurements of the MaNGA Data Analysis Pipeline or due to low H$\alpha$ equivalent widths. This confirms our previously mentioned result of our AGN selection algorithm being biased against gas poor galaxies with weak emission lines. This analysis also shows that the current version of the MaNGA Data Analysis Pipeline is not optimized yet to account for extremely broad emission lines with FWHM~$> 800$~km/s. 

\subsection{Ancillary AGN}

As described in Section 2.4, there are 13 sources in MaNGA that are part of an ancillary program within MaNGA targeting AGN that were selected using a variety of methods (based on their X-ray properties, optical emission-line ratio measurements or mid-IR WISE photometric properties, Table \ref{ancillary}). Nine of these ancillary AGN are also in our sample of final 303 MaNGA AGN candidates. Notably, all [OIII]-selected ancillary sources make it into our sample, as well as the X-ray-selected ancillary AGN and one WISE-selected AGN (MaNGA 1-149211).

We note that MaNGA 1-209980 is an ideal example of a galaxy in which multiple processes contribute to the fact that almost all spaxels are flagged as AGN or LI(N)ER-like. While the center of the galaxy is clearly dominated by AGN-like emission with high H$\alpha$ equivalent widths, the gas signatures at larger radii are more consistent with expectations from the pAGB or diffuse ionized gas scenario. Such kind of sources initially motivated our choice of computing the H$\alpha$ equivalent width using only the top 20\% of the distribution of the spaxel-based equivalent width measurements in the AGN and LI(N)ER classified spaxels, respectively. Computing H$\alpha$ equivalent widths by averaging over all AGN or LI(N)ER classified spaxels in this particular galaxy would have not resulted in a mean equivalent width of $> 5$\AA\ because the contributions from diffuse ionized gas regions and hot stars would have smeared out the signal. 

We furthermore inspect the four WISE-selected ancillary AGN that were not picked up by our MaNGA AGN selection and note that all of them were not selected for different reasons: MaNGA 1-47256 does not pass our cut on the equivalent width of H$\alpha$, MaNGA 1-177270 does not pass our cut in H$\alpha$ surface brightness, the emission-line ratio measurement of MaNGA 1-24423 are very consistent with ionisation through star-formation processes alone and MaNGA 1-90901 has low spaxel fractions f$_{\rm{A,N}}$ and f$_{\rm{AL, S}}$. All in all, none of these four galaxies exhibits convincing optical signatures that would point towards an AGN residing in those galaxies. This does not mean, however, that there is no AGN in these galaxies. It is well known that different AGN selection techniques come with strong biases. On the other hand, the mid-IR AGN selection is known to suffer from significant contamination at low luminosities from starburst galaxies. Although this contamination has been tried to be mitigated by applying a luminosity cut of $L_{bol} > 10^{43}$~erg~s$^{-1}$ in the initial selection of ancillary targets, the contamination may not be negligible. We remind the reader that the comparison in this Section is based on very low number statistics and more Ancillary AGN in MaNGA (particularly WISE-selected ones) will have to be observed to evaluate the overlap and biases of different selection methods. 

\begin{figure}
\centering
\includegraphics[scale = 0.43]{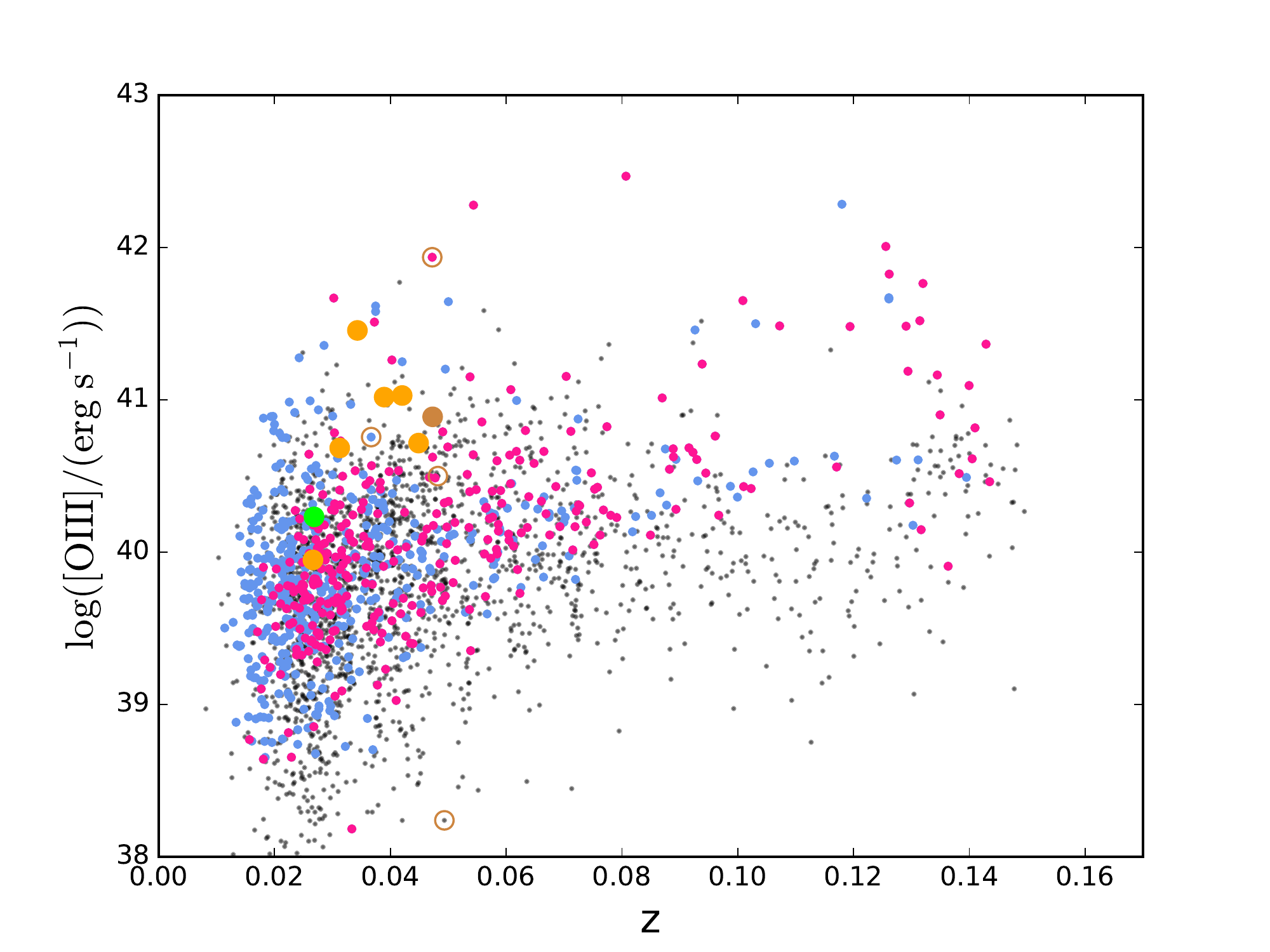}
\caption{[OIII] luminosity vs. redshift for all MaNGA galaxies (small, grey data points), the initial AGN candidates (blue points) and the final AGN candidates (pink points). Additionally, we show where the MaNGA ancillary AGN which were originally classified as AGN using BAT X-ray observations, [OIII] line flux or WISE IR-colors) lie in this space. We show the BAT-selected source in lime green, the [OIII]-selected sources in orange and the WISE-selected sources in brown. The open symbols show the WISE-selected AGN that did not pass the here presented AGN selection. }
\label{anc}
\end{figure}

In Figure \ref{anc}, we show the distribution of all MaNGA galaxies, the initial and final AGN candidates and the ancillary AGN in the [OIII]-$z$ plane. The [OIII] luminosity is a commonly used indicator of AGN bolometric luminosity if an AGN is present in the galaxy \citep{Heckman_2005, Reyes_2008}. Generally, high [OIII] luminosities are indicative of AGN activity in a galaxy by itself \citep{Zakamska_2003, Reyes_2008, Yuan_2016}. This shows that our AGN detection algorithm manages to select likely AGN, i.e. the high [OIII] luminosity sources. The ancillary AGN (shown with large symbols) belong -- partly by design -- to the most [OIII]-luminous objects at each redshift. We note that with the exception of one source, all WISE AGN that did not make our AGN selection (open symbols) also belong to some of the most [OIII] luminous sources at their redshift. This shows that additional AGN are expected among the MaNGA galaxies and that the here presented emission-line based algorithm is not exclusive. In Wylezalek et al. in prep. we will assess the multi-wavelength AGN signatures of all MaNGA galaxies and the here selected AGN candidates.

%
%

\section{Conclusions}

In this paper, we have analyzed all galaxies that have been observed by the optical fibre-bundle IFU survey SDSS-IV MaNGA with respect to their resolved BPT classifications with the goal of developing an IFU based AGN selection. The primary challenge when selecting AGN based on spatially resolved optical IFU observations is how to overcome contaminating processes that mimic AGN-like signatures of photo-ionized gas. The sources of contamination are particularly problematic in galaxy-wide IFU observations, such as the MaNGA survey, since they are a strong function of distance to the galactic center. Known sources of contamination include diffuse ionized gas and extraplanar gas in which changes of the ionization and metallicity lead to enhanced optical line ratios. Furthermore, BPT diagnostics can be contaminated by photo-ionization due to hot evolved stars such as pAGB stars. 

Several methods to circumvent this contamination have been suggested in the literature such as applying additional cuts on H$\alpha$ surface brightness and the H$\alpha$ equivalent width. We analyze MaNGA galaxies in the space of AGN spaxel fraction (in every galaxy, the fraction of spaxels that show AGN-like ionization) vs. H$\alpha$ equivalent width (right panel in Figure \ref{fractions}). In this space, we successfully recover the well-known bi-modality of local galaxies. Specifically, we find that there are two distinct populations of galaxies. One with low AGN+LI(N)ER spaxel fraction f$_{\rm{AGN+LIER, [SII]}}$ and high H$\alpha$ equivalent width which we identify with gas-rich, star-forming galaxies. The other population presents with a high AGN+LI(N)ER spaxel fraction f$_{\rm{AGN+LIER, [SII]}}$ and low H$\alpha$ equivalent width, as well as a high stellar mass. This population is naturally identified with gas-poor massive galaxies.

AGN tend to occupy the transitional region in the space of AGN-like spaxel fraction and H$\alpha$ equivalent width. Cuts on H$\alpha$ equivalent width previously proposed in the literature successfully clean the sample of particularly high mass, low sSFR galaxies that are dominated by old stellar populations and whose high ionization is likely due to hot old stars. These additional cuts on H$\alpha$ surface brightness and the H$\alpha$ equivalent width, however, only address the problem of distinguishing between AGN and non-AGN in the high f$_{\rm{AGN+LIER, [SII]}}$ regime. 

Even after we apply the suggested additional cuts on H$\alpha$ surface brightness and the H$\alpha$ equivalent width, a large fraction of sources displays a significant amount of AGN-like emission in objects that are unlikely to host an AGN. This is especially true for low mass, high sSFR galaxies which may show large regions of AGN or LI(N)ER-like emission and pass all cuts in H$\alpha$ surface brightness and in H$\alpha$ equivalent width. A simple inspection of [SII]-BPT maps of such galaxies may lead to the impression that they host AGN. Closely inspecting the distribution of line-ratios, however, reveals that the majority of the AGN or LI(N)ER-like classified spaxels lie very close to the demarcation line in the BPT diagram. Such spaxels therefore rather represent the high ionization signature tail of the distribution of the star forming-spaxels than truly enhanced line ratios that would point towards a harder ionization source (such as an AGN). In addition to added contamination from diffuse ionized gas regions and old, hot stars, contributions from young, massive, hot stars such as Wolf-Rayet stars can also lead to enhanced line ratios. This is especially likely in low mass, blue galaxies with high sSFR \citep{Brinchmann_2008}. 

We have developed a method to account for these effects by measuring the distance of the AGN+LI(N)ER spaxels from the star formation demarcation line in the [SII]-BPT. Computing the mean distance $d_{BPT}$ of the 20\% of the spaxels with the largest distances provides us with a quantitative measurement for the significance of the BPT classifications. The distribution of $d_{BPT}$ for initial AGN candidates is not bi-modal. This would have allowed for an easy distinction between the highly sSFR galaxies contaminating our AGN selection and the high-significance AGN candidates. By visually inspecting many galaxies with low $d_{BPT}$, we have adopted $d_{BPT} > 0.3$ as an additional cut to our AGN selection. This choice is also driven by the fact that one of the first MaNGA AGN candidates that was followed up with higher spatial resolution IFU observation with Gemini-GMOS \citep[Blob Source, ][]{Wylezalek_2017} has $d_{BPT} = 0.35$, revealing that there was indeed a hidden AGN in the center. Such weakly ionizing, potentially young AGN with small outflows are important for understanding detailed AGN feeding and feedback processes. The challenge for large IFU surveys therefore lies in disentangling such weakly ionizing AGN from the population of low mass, high sSFR galaxies. 

In addition to accounting for hot, old stars through cuts on H$\alpha$ equivalent width, we therefore advocate for and apply an additional cut in $d_{BPT}$ to account for uncertainty in the star-formation demarcation line and contamination from hot, young stars (such as Wolf-Rayet stars). This refined AGN selection leaves us with 303 AGN candidates out of 2778 sources in the current MaNGA data release. By cross-matching with emission line flux catalogs for the SDSS single-fibre observations we investigate how our AGN candidates would have been classified based on single-fibre observations alone. AGN selection based on detecting and measuring optical emission lines naturally biases against high mass, gas poor galaxies with weak emission lines. Many of such sources may be missed in the MaNGA-based AGN selection, although they may previously have been correctly identified to be AGN based on the single-fibre spectra. 

As expected, our initial AGN selection (before imposing the cut on $d_{BPT}$) included many galaxies that had been clearly classified as star-forming based on single-fibre spectra. Imposing the additional cut on $d_{BPT}$ decreases the contamination by these high sSFR galaxies significantly. But even after we impose the additional cut on $d_{BPT}$, our AGN selection includes many sources which would have been classified as star-forming galaxies based on single-fibre observations. We show two such sources in the middle and bottom row of Figure \ref{AGN_selection_final}. These sources show unambiguous signatures of photo-ionization by an AGN beyond $3\arcsec$. These examples highlight the power of large IFU surveys: not only do they allow to uncover AGN which might be hidden behind large column densities of dust at the center of their host galaxy through their signatures at larger distances, but they also allow to uncover AGN that might have `just' turned off and do not show signatures close to the galactic center anymore but where relic AGN photo-ionization signatures, so-called light echoes, can still be detected. Additionally, IFU surveys such as MaNGA bear the potential to uncover off-nuclear AGN after a recent or ongoing galaxy merger. Having access to this completely new parameter space in finding and characterizing AGN may have important constraints on galaxy evolution models and the galaxy-black hole co-evolution. In a forthcoming paper (Wylezalek et al. in prep.) we will carry out an extensive multi-wavelength analysis of the AGN candidates in MaNGA to investigate their true nature and their host galaxy properties.

\section*{Acknowledgements}
D.W. acknowledges support by the Akbari-Mack Postdoctoral Fellowship and the JHU Provost's Postdoctoral Diversity Fellowship. 

We thank the Lorentz Center in Leiden that hosted the workshop `Observations and Theory of Quasar Outflows' during which many fruitful discussions took place. In particular, we thank Michael Eracleous for useful comments. 

This research made use of Marvin, a core Python package and web framework for MaNGA data, developed by Brian Cherinka, Jos\'{e} S\'{a}nchez-Gallego, and Brett Andrews. (MaNGA Collaboration, 2017).

Funding for the Sloan Digital Sky Survey IV has been provided by
the Alfred P. Sloan Foundation, the U.S. Department of Energy Office of
Science, and the Participating Institutions. SDSS-IV acknowledges
support and resources from the Center for High-Performance Computing at
the University of Utah. The SDSS web site is www.sdss.org.

SDSS-IV is managed by the Astrophysical Research Consortium for the 
Participating Institutions of the SDSS Collaboration including the 
Brazilian Participation Group, the Carnegie Institution for Science, 
Carnegie Mellon University, the Chilean Participation Group, the French Participation Group, Harvard-Smithsonian Center for Astrophysics, 
Instituto de Astrof\'isica de Canarias, The Johns Hopkins University, 
Kavli Institute for the Physics and Mathematics of the Universe (IPMU) / 
University of Tokyo, Lawrence Berkeley National Laboratory, 
Leibniz Institut f\"ur Astrophysik Potsdam (AIP),  
Max-Planck-Institut f\"ur Astronomie (MPIA Heidelberg), 
Max-Planck-Institut f\"ur Astrophysik (MPA Garching), 
Max-Planck-Institut f\"ur Extraterrestrische Physik (MPE), 
National Astronomical Observatory of China, New Mexico State University, 
New York University, University of Notre Dame, 
Observat\'ario Nacional / MCTI, The Ohio State University, 
Pennsylvania State University, Shanghai Astronomical Observatory, 
United Kingdom Participation Group,
Universidad Nacional Aut\'onoma de M\'exico, University of Arizona, 
University of Colorado Boulder, University of Oxford, University of Portsmouth, 
University of Utah, University of Virginia, University of Washington, University of Wisconsin, 
Vanderbilt University, and Yale University.




\bibliographystyle{mnras}
\bibliography{/Users/dominika/JHU_Postdoc/Papers_FA/master_bib} 

\bsp	
\label{lastpage}
\end{document}